\documentclass[a4paper,11pt]{article}
\pdfoutput=1 

\usepackage{jheppub} 
\usepackage{amsmath,amssymb,amsthm,amscd,graphicx}
\input epsf.sty

\theoremstyle{definition}

\newcommand{\CB}{{\cal B}}
\newcommand{\CC}{{\cal C}}

\newcommand{\CI}{{\cal I}}

\newcommand{\CK}{{\cal K}}
\newcommand{\CL}{{\cal L}}

\newcommand{\CO}{{\cal O}}

\def\IR{{\mathbb R}}

\newcommand{\mD}{\mathsf{D}}
\newcommand{\mO}{\mathsf{O}}

\newcommand{\mH}{\mathsf{H}}

\newcommand{\bk}{{\bf k}}
\newcommand{\bq}{{\bf q}}

\newcommand{\bP}{{\bf P}}

\newcommand{\re}{{\rm e}}
\newcommand{\ri}{{\rm i}}
\newcommand{\rd}{{\rm d}}

\newcommand{\be}{\begin{equation}}
\newcommand{\ee}{\end{equation}}
\newcommand{\ba}{\begin{aligned}}
\newcommand{\ea}{\end{aligned}}
\newcommand{\ben}{\begin{eqnarray}\displaystyle}
\newcommand{\een}{\end{eqnarray}}

\newcommand{\sectiono}[1]{\section{#1}\setcounter{equation}{0}}
\renewcommand{\theequation}{\thesection.\arabic{equation}}

\newdimen\tableauside\tableauside=1.0ex
\newdimen\tableaurule\tableaurule=0.4pt
\newdimen\tableaustep
\def\phantomhrule#1{\hbox{\vbox to0pt{\hrule height\tableaurule width#1\vss}}}
\def\phantomvrule#1{\vbox{\hbox to0pt{\vrule width\tableaurule height#1\hss}}}
\def\sqr{\vbox{%
  \phantomhrule\tableaustep
  \hbox{\phantomvrule\tableaustep\kern\tableaustep\phantomvrule\tableaustep}%
  \hbox{\vbox{\phantomhrule\tableauside}\kern-\tableaurule}}}
\def\squares#1{\hbox{\count0=#1\noindent\loop\sqr
  \advance\count0 by-1 \ifnum\count0>0\repeat}}
\def\tableau#1{\vcenter{\offinterlineskip
  \tableaustep=\tableauside\advance\tableaustep by-\tableaurule
  \kern\normallineskip\hbox
    {\kern\normallineskip\vbox
      {\gettableau#1 0 }%
     \kern\normallineskip\kern\tableaurule}%
  \kern\normallineskip\kern\tableaurule}}
\def\gettableau#1{\ifnum#1=0\let\next=\null\else
\squares{#1}\let\next=\gettableau\fi\next}

\newcommand{\gyl}{\log\left(\frac{\pi z}{4}\right)}
\newcommand{\figref}[1]{Fig.~\protect\ref{#1}}

\title{\huge{\boldmath Resurgence for superconductors}}

\author{Marcos Mari\~no and Tom\'as Reis}

\affiliation{D\'epartement de Physique Th\'eorique et Section de Math\'ematiques\\
Universit\'e de Gen\`eve, Gen\`eve, CH-1211 Switzerland}

\abstract{An important non-perturbative effect in quantum physics is the 
energy gap of superconductors, which is exponentially small in the coupling constant. 
A natural question is whether this effect can be incorporated in the theory of resurgence. 
In this paper we take some steps in this direction. We conjecture that the perturbative series for the ground state energy of a superconductor 
is factorially divergent, and that its leading Borel singularity is governed by the superconducting energy gap. 
We test this conjecture in detail in the attractive Gaudin-Yang model, an exactly solvable model in one dimension with a BCS-like ground state. 
In order to do this, we develop techniques to calculate the exact perturbative series of its 
ground state energy up to high order. We also argue that the Borel singularity is of the renormalon type, and we identify a class of diagrams leading to factorial growth. 
We give additional evidence for the conjecture in other models. }    

\begin{document}
\maketitle
\flushbottom
 
\sectiono{Introduction}

The energy gap of a superconductor is a robust property of Fermi systems with a weak attraction. It has a non-analytic, exponentially small dependence on the coupling constant \cite{bcs}, and 
it should be therefore regarded as a non-perturbative effect. It is well known that many non-perturbative effects in physics, from quantum mechanics to string theory, 
can be detected by looking at the large-order behavior of the perturbative series. In fact, they provide a detailed characterization of the perturbative series at large orders, as first 
shown in the pioneering work of Bender and Wu \cite{bw,bw2}. This relationship between perturbative series and non-perturbative effects 
has evolved into the theory of resurgent asymptotics, which can be considered as a general framework to understand 
non-perturbative phenomena in physics and mathematics (see \cite{mmlargen,abs} for reviews and references). 

It is then natural to ask whether the theory of resurgence has something to say about the energy gap of a superconductor, providing in this way a 
characterization of the behavior of the perturbative series  
for weakly interacting fermion systems. In this work we take some steps in this direction. We argue that the perturbative series for the ground 
state energy of a superconductor is factorially divergent and non-Borel 
summable, and that the corresponding non-perturbative ambiguity is closely related the energy gap. More precisely, let 
\be
\label{en-gap}
\Delta^2 \sim \re^{-A/\gamma} 
\ee
be the square of the superconducting energy gap, where $\gamma$ is the coupling constant. Then, we conjecture that the first singularity in the Borel plane of the coupling 
constant is given by $A$. The fact that the non-perturbative ambiguity involves $\Delta^2$ can be justified heuristically by noting that in BCS theory the non-analytic correction 
to the ground state energy is proportional to $\Delta^2$. 

Some aspects of our proposal have been hinted before in the literature, although the behavior of perturbative series in many-body theory has not been 
studied extensively. Previous works pointed out the possibility that the series diverges factorially \cite{baker-review,jw}, although instanton estimates in 
Fermi systems have suggested a milder growth of the coefficients \cite{parisi-fermi,baker-pirner}, due to cancellations among diagrams. The Cooper instability has been sometimes linked to the lack 
of analyticity of the perturbative series, but no precise connection seems to have been made 
between these two phenomena. One possible reason for the scarcity of studies of large order behavior in many-body theory is probably the lack of examples where 
the perturbative series can be calculated to sufficiently large order. In Quantum Mechanics, calculable models  
have been crucial in the development of the theory of resurgence. For example, the 
groundbreaking results of Bender and Wu in \cite{bw,bw2} were made possible to a large extent by the data provided by the perturbative series for the ground state energy of the quartic oscillator. 
 
In order to make precise statements about the perturbative series in many-body fermionic systems, it is 
useful to have an example akin to the anharmonic oscillator in quantum mechanics, where the ground 
state energy can be explicitly computed to high order in perturbation theory. A natural candidate is the Gaudin--Yang model in 
one dimension \cite{gaudin, yang}. The ground state of the Gaudin--Yang model is a BCS-like state, 
and there is a superconducting gap which is non-perturbative 
in the coupling constant \cite{ko} (see \cite{giamarchi} for this and other results on one-dimensional models). At the same time, the energy of the ground state can be computed exactly with the Bethe 
ansatz, through a simple integral equation. 

However, extracting the perturbative series from this integral equation turns out to 
be highly non-trivial. The same difficulty was observed in the closely related Lieb--Liniger model \cite{ll}. 
In spite of many efforts \cite{iw-gy,tw1,tw2, guan-review}, and after fifty years, only the first three terms of 
the series have been obtained analytically (further coefficients have been obtained numerically in \cite{prolhac}). In this paper, 
building upon previous work \cite{hutson, popov,volin,volin-thesis}, we develop a method to calculate analytically the perturbative series from the Bethe ansatz 
equations, and we obtain explicit results for its coefficients up to order $50$ in the coupling constant (in principle, one can go further by using enough CPU time). 
This allows us to study the large order behavior of the series and to determine (numerically) the 
leading non-perturbative ambiguity. We find that the first singularity in the Borel plane corresponds indeed to the 
energy gap, in agreement with the conjecture above. 

In general, there are two different sources for the factorial behavior of a perturbative series (see \cite{mmbook} for a review): 
instanton singularities and renormalon singularities. Instanton singularities are due to non-trivial saddles in the path integral, and 
they encode the factorial growth in the number of diagrams. Renormalon singularities are typically associated to sequences of diagrams in 
which each diagram diverges factorially with the loop order, after integration over momenta (see e.g. \cite{beneke} for a review and references). 
We argue that the large order behavior found in the Gaudin--Yang model should be attributed to renormalon-type singularities. To support this, 
we show explicitly that both ring and ladder diagrams lead to factorially divergent sequences. 

We give further evidence for our conjecture by examining an 
even simpler example, the one-dimensional Hubbard model at half-filling \cite{lw}. 
Here the series for the ground state energy is known in closed form, and we verify our conjecture analytically. We also examine the much more complicated case of the dilute Fermi gas in three dimensions. 
Although a systematic calculation of the ground state energy perturbative series at large order is currently out of reach, we find some encouraging indications that it might have the 
appropriate renormalon-type singularities. 

This paper is organized as follows. In section \ref{gy-intro} we present various approaches to the study of the ground state energy density of the Gaudin--Yang model: the perturbative approach, the BCS approach, and finally the exact Bethe ansatz solution. Most of section \ref{gy-intro} is a review of known facts, but some of the material included here, like the explicit perturbative results for the ring and ladder diagrams, seem to be new. In section \ref{sec-pert-series} we present in detail the method to extract the perturbative series from the Bethe ansatz equation. 
We apply it to the Gaudin--Yang model in the attractive and repulsive regimes, and to its generalization for $\kappa$ spin components. 
In section \ref{lo-sec} we study the large order behavior of the series, we 
make the connection to the superconducting energy gap, and we argue that the Borel singularity is of the renormalon type. In section \ref{sec-models} we consider 
the one-dimensional Hubbard model and the dilute Fermi gas in three dimensions in the light of our conjecture. Finally, in section \ref{sec-conclude}, we conclude and and list many open problems. There are also two Appendices. In the first one we evaluate the NNNLO diagram (of order $\gamma^3$) in the Gaudin--Yang model. In the second one we give some details of the matching procedure explained in section \ref{sec-pert-series}. 

\sectiono{The Gaudin--Yang model}
\label{gy-intro}

\subsection{The perturbative approach}
\label{pert-subsec}

\begin{figure}[!ht]
\leavevmode
\begin{center}
\includegraphics[height=2cm]{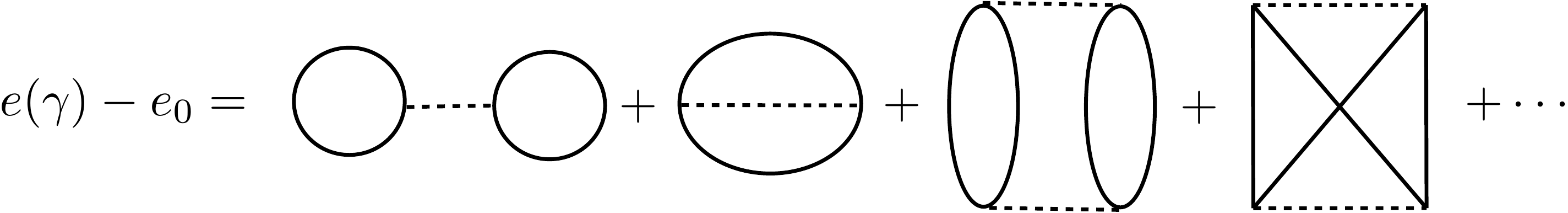}
\end{center}
\caption{Feynman diagrams contributing to the ground state energy density at the very first orders. The first and second diagram are respectively the Hartree and Fock contributions.}
\label{pert-series}
\end{figure}

The Gaudin--Yang model describes a non-relativistic, one-dimensional gas of $N$ spin $1/2$ fermions with $\delta$-function interactions. The potential is given by 
\be
\label{potential}
V(x)= -2 c \delta(x).
\ee
 The Hamiltonian of the system is
\be
\label{h-gy}
H_N=-\sum_{j=1}^N {\partial^2 \over \partial x_j^2}- 2 c \sum_{ 1\le i<j\le N} \delta(x_i-x_j), 
\ee
i.e. we set $\hbar=2m=1$. We will mostly consider the attractive case. In our conventions, this corresponds to a positive coupling constant: $c>0$. Let $E_0(N, L)$ be the ground state energy for the 
$N$-particle system on a circle of length $L$. In the thermodynamic limit 
of fixed density $n$, 
\be
N \rightarrow \infty, \qquad L \rightarrow \infty, \qquad n={N\over L} \, \, \, \text{fixed}, 
\ee
we can define the ground state energy density as 
\be
E= \lim_{L \to \infty} {E_0(N, L) \over L}.
\ee
 The Fermi momentum $k_F$ is related to the density by
\be
k_F= {\pi n \over 2}. 
\ee
Since $c$ has the dimension of an inverse length, the dimensionless coupling constant is 
\be
\label{gamcn}
\gamma={c \over n}. 
\ee
Note that weak coupling, $\gamma \ll1$, corresponds here to {\it high} density, in contrast to what happens in the three dimensional Fermi gas 
with a delta function interaction. The ground state energy density can be written as a formal power series in $\gamma$ of the form 
\be
\label{e-cos}
e(\gamma)=e_0+ \sum_{n \ge  1} c_n \gamma^n, 
\ee
where
\be
\label{egamma}
e(\gamma)={E(\gamma) \over n^3} 
\ee
and
\be
e_0= {\pi^2 \over 12}
\ee
corresponds to the free Fermi gas. The coefficients $c_n$, $n \ge 1$, can be calculated in terms of Feynman diagrams. 
The coefficient $c_1$ is given by the sum of Hartree and Fock terms, and its calculation is an elementary exercise:
\be
c_1=- {1\over 2}. 
\ee
The coefficient $c_2$ is obtained by the sum of the two-bubble diagram in \figref{pert-series} and the diagram obtained by exchange from it. It has been calculated in \cite{magyar}, and its value is 
\be
c_2=-{1\over 12}. 
\ee
The next coefficient was computed in \cite{magyar2}, and its value is:
\be
\label{c3}
c_3=-{\zeta(3) \over \pi^4}. 
\ee
In Appendix \ref{3-diagrams} we rederive this result for an arbitrary number of spin components. 
Clearly, pushing this calculation to higher orders becomes increasingly hard, and as far as we know, no other results for higher coefficients have been 
obtained in the literature.

It is however possible to consider partial resummations of the perturbative series involving special types of diagrams. 
The most popular ones are the ring diagrams appearing in the RPA, and the ladder diagrams. In particular, one can 
obtain interesting weak-strong coupling interpolations for Fermi gases in three dimensions by resumming these diagrams 
\cite{steele,hf, schafer, kaiser1, kaiser2, kaiser2d}. 

Let us then start by considering the resummation of ring diagrams, i.e. the random phase approximation (RPA). The building block of the RPA approximation is the 
so-called polarization loop or Lindhard function, which is given in $d$ dimensions by
\be
\label{polar}
\Pi(q)=-\ri \int_{\IR^{d+1}}{\rd^{d+1} k \over (2 \pi)^{d+1}} G(k+q) G(k). 
\ee
Here, $G(k)$ is the free fermion propagator at $T=0$. In terms of the Lindhard function, the contribution of a ring diagram with $\ell \ge 2$ bubbles to the ground state energy density is 
\be
\label{ell-in}
{\ri \over 2 \ell} \int_{\IR^{d+1}} {\rd^{d+1}q  \over (2 \pi)^{d+1}} \left( \kappa \Pi(q) \widehat V (\bq) \right)^\ell, 
\ee
where $\kappa$ is the number of spin components and $\widehat V(\bq)$ is the Fourier transform of the potential. By rotating the integration contour and taking 
into account that $\Pi(\bq, \ri \omega)$ is an even function of $\omega$, we can write (\ref{ell-in}) as 
\be
\label{l-ring}
-{1\over \ell} \int_{\IR^{d}} {\rd^d \bq  \over (2 \pi)^{d}} \int_0^\infty {\rd \omega \over 2 \pi}  \left( \kappa \Pi(\bq, \ri \omega) \widehat V (\bq) \right)^\ell. 
\ee
It is possible to sum these terms in closed form to find the usual RPA contribution to the correlation energy, 
\be
\label{log-ring}
E_{\rm RPA}= 
\int_{\IR^{d}} {\rd^d \bq  \over (2 \pi)^{d}} \int_0^\infty {\rd \omega \over 2 \pi}\left\{  \log\left(1-\kappa \Pi(\bq, \ri \omega) \widehat V (\bq) \right)+ \kappa \Pi(\bq, \ri \omega) \widehat V (\bq) \right\}. 
\ee
We note that in one dimension, the Lindhard function can be written as \cite{electron-liquid}
\be
 \Pi(q, \ri \omega)= -{m\over 2 \pi  k_F  y} \log \left( {(y/2+1)^2 + \nu^2 \over (y/2-1)^2 + \nu^2 }  \right),  
 \ee
where the variables $\nu$, $y$ are defined as 
\be
 y={q \over k_F}, \qquad \nu= {m \omega \over q k_F}. 
 \ee

 One important property of ring diagrams is that they give the first non-trivial contribution to the ground state energy density 
 in the limit of a large number $\kappa$ of spin components (see e.g. \cite{hf}). In this limit, which is akin to a large $N$ 
 limit, one has to scale the coupling constant in the potential and the density $n$ in such a way that every interaction gives a power of 
 $\kappa^{-1}$ and that the Fermi momentum is fixed. 
 
\begin{figure}[!ht]
\leavevmode
\begin{center}
\includegraphics[height=3.25cm]{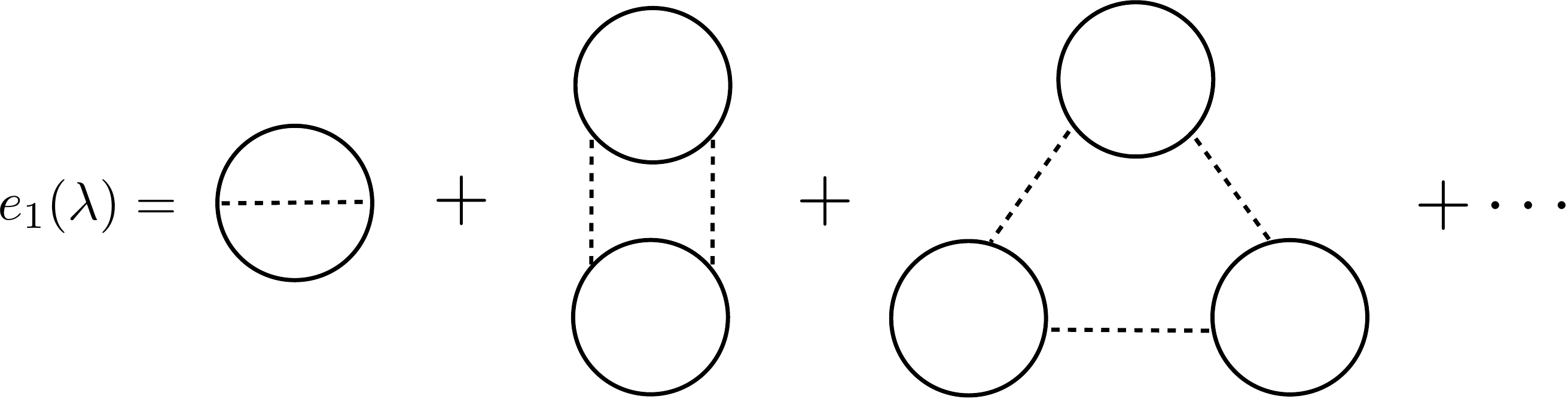}
\end{center}
\caption{The next-to-leading correction to the ground state energy density in the large $\kappa$ limit is given by the sum of ring diagrams.}
\label{rings}
\end{figure}

 Let us now analyze the ring diagrams and the large $\kappa$ limit 
 in the case of the Gaudin--Yang model. For a delta function potential (\ref{potential}) $\widehat V(\bq)=-2c$. 
 The Fermi momentum is given by $k_F= \pi n/\kappa$, therefore the appropriate scaled coupling constant is
\be
\label{lam-thooft}
\lambda=  \left( {\kappa \over 2} \right)^2 \gamma. 
\ee
It plays the r\^ole of a 't Hooft parameter and is kept fixed in the large $\kappa$ limit. We also define the rescaled ground energy density as 
\be
\label{elambda}
e(\lambda; \kappa)= {1\over 4} {E/\kappa \over (n/\kappa)^3}.
\ee
It has the following large $\kappa$ expansion, 
\be
\label{large-kappa}
e(\lambda; \kappa)= \sum_{n=0}^\infty e_n(\lambda) \kappa^{-n}. 
\ee
Note that, when $\kappa=2$, $\lambda$ and $e (\lambda; \kappa)$ become the coupling constant 
$\gamma$ and the energy density $e(\gamma)$, respectively. The leading order function $e_0(\lambda)$ to (\ref{large-kappa}) 
is given by the free gas result plus the Hartree term, 
\be
e_0(\lambda)={\pi^2 \over 12} - \lambda. 
\ee
The subleading function $e_1(\lambda)$ is the sum of the contributions of ring diagrams, 
\be
\label{sum-ring}
e_1(\lambda)= \lambda - {\pi \over 4} \sum_{\ell=2}^\infty  {1\over \ell}\left( {4 \lambda \over \pi^2} \right)^\ell \CI_\ell, 
\ee
where the first term corresponds to the Fock diagram, and 
\be
\label{il-int}
\CI_\ell =\int_0^\infty \rd y \, y \int_0^\infty \rd \nu  \left( {1\over 2 y} \log \left( {(y/2+1)^2 + \nu^2 \over (y/2-1)^2 + \nu^2 }  \right) \right)^\ell. 
\ee
We can represent this result schematically as in \figref{rings}.

\begin{figure}[!ht]
\leavevmode
\begin{center}
\includegraphics[height=2.5cm]{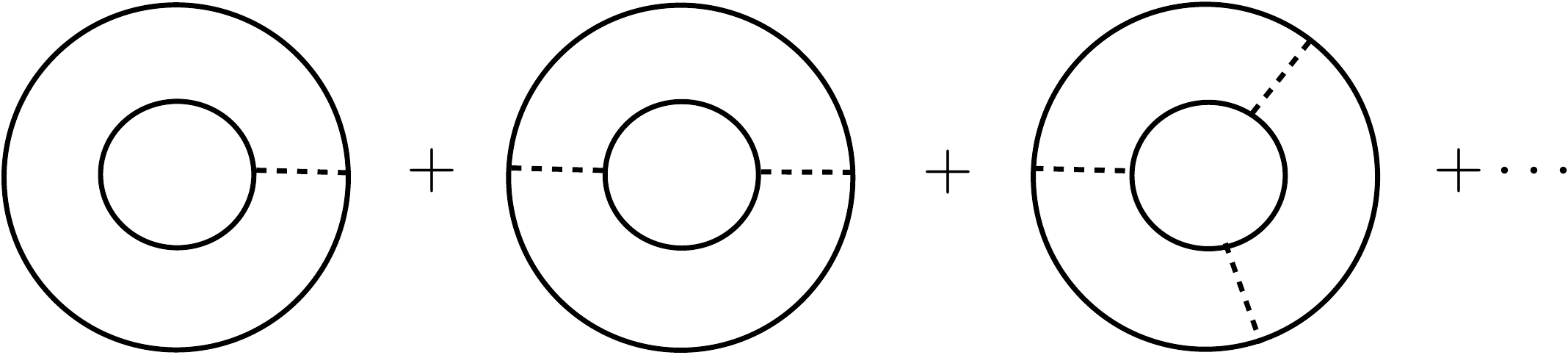}
\end{center}
\caption{Ladder diagrams.}
\label{ladders}
\end{figure}

Another useful reorganization of the perturbative series is the particle-particle ladder expansion. In this approach, one considers diagrams with the shape of a ladder with $n$ steps, where each of the steps is an interaction. The ladder is then closed into a circle, as shown in \figref{ladders}, in order to have a vacuum diagram. We will follow the approach in 
\cite{steele, schafer,he-huang} and consider ladders in which two of the parallel lines are ``hole" lines (i.e. their momenta are inside the Fermi sphere), while the 
rest of the lines are ``particle" lines 
(i.e. with momenta outside the Fermi sphere). We will refer to the resulting diagrams as pp-ladders (or particle-particle ladders), although a 
more appropriate name would be $n$pp-$1$hh bubbles \cite{he-huang}. 
We will also include the diagrams obtained from the ladders by exchange of the hole lines. 

Let us summarize the calculation of pp ladder diagrams, following the presentation in \cite{he-huang}. 
The building block of the pp ladder approximation is the particle-particle bubble in $d$ dimensions, 
\be
\Pi_{\rm pp}(P)= \ri \int_{\IR^{d+1}}{\rd^{d+1} q \over (2 \pi)^{d+1}} G_{\rm p}(P/2-q) G_{\rm p}(P/2+q),  
\ee
where $G_{\rm p}(k)$ is the particle part of the propagator (in which $|\bk|>k_F$). In the calculation of the ground state energy, one needs the on-shell version of $\Pi_{\rm pp}(P)$, which is given by 
\be
\Pi_{\rm pp} ({\bf P},{\bf k})=m \int_{\IR^{d}}\frac{\rd^{d}{\bf q}}{(2\pi)^{d}}\frac{\theta(|{\bf P}/2+{\bf
q}|-k_{\text F})\theta(|{\bf P}/2-{\bf q}|-k_{\text
F})}{{\bf k}^2-{\bf q}^2+\ri\epsilon}. 
\ee
Here $\theta(x)$ is the Heaviside function. For a delta function interaction (\ref{potential}), the contribution of pp ladders to the ground state energy is a geometric series which can be resummed as 
 \be
-2c \int_{\IR^d}{\rd^d \bP\over (2\pi)^d}\int_{\IR^d}\frac{\rd^d {\bf
k}}{(2\pi)^d}\frac{\theta(k_{\text F}-|\bP/2+\bk|)\theta(k_{\text F}-|\bP/2-\bk|)}{1+2c \Pi_{\rm pp} ({\bf
P},{\bf k})}.
\ee
Note that only the real part of $\Pi_{\rm pp} ({\bP},{\bf k})$ contributes to this integral. We introduce now the dimensionless variables 
\be
s={|\bP| \over 2k_F}, \qquad t= {|\bk| \over k_F}. 
\ee
Specializing now to one dimension, we find
 \be
{\rm Re}\, \Pi_{\rm pp}(P,k)= {m \over 2 \pi  k_F} F_{\rm pp}(s, t), 
 \ee
 where
 \be
 \label{pp1}
 F_{\rm pp}(s, t)= {1\over  t} \log\left( {1+ s-t \over 1+s + t} \right). 
 \ee
We conclude that the pp ladder contribution to $e(\gamma)$, which we will denote by $e_{\rm pp}(\gamma)$, is given by
\be
\label{e-pp-ladder}
e_{\rm pp}(\gamma)=e_0 -\gamma \int_0^1 \rd s \int_0^{1-s} \rd t {1\over 1+ \gamma F_{\rm pp}(s,t)/\pi^2}. 
\ee
This can be re-expanded in power series of $\gamma$ to obtain, 
\be
\label{epp-coefs}
 e_{\rm pp} (\gamma)=\sum_{n \ge 0} c_n^{\rm pp} \gamma^n, 
\ee
where the very first coefficients can be computed explicitly in closed form, 
\be
\label{cpp-values}
c^{\rm pp}_1=-{1\over 2}, \qquad 
c^{\rm pp}_2=-{1\over 12}, \qquad 
c^{\rm pp}_3= {9 \zeta(3) \over 2 \pi^4}- {\log(2) \over \pi^2}.
\ee
The first two coefficients correspond to the diagrams shown in \figref{pert-series}, i.e. to the Hartree--Fock contributions and to the contributions at order $\gamma^2$. 
Note that the perturbative series up to order $\gamma^2$ is correctly reproduced by the ladder resummation.

\subsection{The BCS approach}

Perturbation theory has of course many limitations, in particular if one is interested in the strong coupling regime where $\gamma \gg 1$. Resummations 
of particular types of diagrams can be sometimes extrapolated to this regime \cite{schafer,kaiser1}, but it is important to have other approaches to the problem. One possibility is 
to use BCS theory. Indeed, in the attractive case, the Hamiltonian (\ref{h-gy}) leads to Cooper pairing, and the ground state of the Gaudin--Yang model can be regarded as a superconductor. 
BCS theory leads to approximate expressions for the energy gap and for the ground state energy which interpolate between 
weak and strong coupling. The application of BCS theory to 
the Gaudin--Yang model was developed in \cite{montse,quick}, and we will complement the results in those papers by presenting explicit expressions for all quantities in 
terms of elliptic functions, as it was done in \cite{strinati} in the three-dimensional case. 

In the BCS approach, we assume that the ground-state wavefunction is of the BCS type and we use the variational or mean-field method (a detailed exposition can be found in section 37 of \cite{fw}). To write 
down the resulting equations, we introduce 
\be
\xi_k=k^2 - b, \qquad b=\mu+ cn, 
\ee
where $\mu$ is the chemical potential and $n$ is the density. The variational method leads to the famous gap equation for the energy gap $\Delta_{\rm BCS}$, which in this case reads
\be 
{c \over 2 \pi} \int_\IR {\rd k \over {\sqrt{\xi_k^2 + \Delta_{\rm BCS}^2}}}=1. 
\ee
In addition, we have to relate $\mu$ to the density, and this gives the equation
\be
n={1\over 2 \pi} \int_\IR \rd k \left( 1- {\xi_k \over {\sqrt{\xi_k^2 + \Delta_{\rm BCS}^2}}} \right). 
\ee
The BCS expression for the ground state energy per particle is 
\be
{E_{\rm BCS} \over n}= {1\over 2 \pi n} \int_\IR \rd k \, \xi_k \left( 1 - {\xi_k \over {\sqrt{\xi_k^2 + \Delta_{\rm BCS}^2}}} \right) + {c n \over 2} +\mu-{\Delta_{\rm BCS}^2 \over 2c n}. 
\ee
We note that this energy incorporates the Hartree--Fock contribution, which is sometimes included only in the fluctuations around the mean-field solution.

It turns out that all the integrals appearing above can be calculated in closed form in terms of elliptic integrals of the first and the second kind. To write explicit expressions, we introduce 
the variable 
\be
\alpha^4=\Delta_{\rm BCS}^2+b^2, 
\ee
as well as the square elliptic modulus 
\be
\label{modulus}
m= {1\over 2} \left( 1+ {1\over {\sqrt{1+ \Delta_{\rm BCS}^2/b^2}}} \right). 
\ee
The gap equation can now be written as 
\be
\label{K-gap}
K(m)= {\pi  \alpha \over c}, 
\ee
where $K(m)$ is the elliptic integral of the first kind. The equation relating the density to $\mu$ is
\be
n= {b \over c} -{ \alpha^2 \over c}+{2 \alpha \over \pi} E(m), 
\ee
where $E(m)$ is the elliptic integral of the second kind. The energy per particle reads  
\be
{E_{\rm BCS} \over n}= \frac{2 \alpha^3}{3 \pi n} \left((2m-1)E(m)+(1-m)\frac{\pi\alpha}{c}\right) -\frac{cn}{2}-\frac{\Delta_{\rm BCS}^2}{2cn}. 
\ee
These expressions are valid for any value of the coupling $c$, but it interesting to analyze them in the weak coupling limit, 
where the gap is exponentially small. It follows from (\ref{modulus}) that in this regime $m \approx 1$. Expanding the functions above around this limit, 
we find an explicit expression for the gap, 
\be
\label{bcs-gap}
\Delta_{\rm BCS} \approx 8 E_F \, \re^{-{\pi^2 \over2 \gamma}}, 
\ee
where $E_F= k_F^2$ is the Fermi energy. Let us note that this is not the correct expression for the energy gap as $\gamma\rightarrow 0$, but rather the approximation 
to it provided by BCS theory. However, it captures already the correct exponentially small dependence on the coupling constant, as we will see. The ground state energy is given by 
\be
\label{ebcs-ex}
{E_{\rm BCS} \over n} \approx {E_0 \over n} -{cn \over 2}-{3\over 8}  {E_0 \over n} \left( {\Delta_{\rm BCS} \over E_F} \right)^2, 
\ee
where $E_0$ is the energy of the free Fermi gas. In terms of $e_{\rm BCS}=E_{\rm BCS}/n^3$, we find 
\be
e_{\rm BCS}(\gamma) \approx {\pi^2\over 12}-{\gamma \over 2} - 2 \pi^2 \re^{- \pi^2 / \gamma}. 
\ee
This expression includes the Hartree--Fock energy, together with exponentially small corrections in $\gamma$. We note that the non-analytic correction is proportional 
to the {\it square} of the gap, $\Delta_{\rm BCS}^2$.

\subsection{The Bethe ansatz approach}

One of the most important aspects of the Gaudin--Yang model is that its ground state energy can be computed exactly by using the Bethe ansatz. This exact 
solution was obtained in \cite{gaudin, yang} for fermions of spin $1/2$, and extended in \cite{taka} to fermions with $\kappa$ spin components. An excellent presentation of the 
spin $1/2$ case can be found in \cite{sb-book}. We will now summarize some ingredients of the solution 
which will be useful in the following. 

We will first consider the attractive case. The basic ingredient is the density of Bethe roots $\phi(k)$ describing the ground state (which has zero polarization). This density has a compact support $ [-K, K]$ and
 satisfies the so-called {\it Gaudin's integral equation}
\be
 \phi(k)+ {c\over \pi}  \int_{-K}^K  { \phi(k') \over  c^2 + (k-k')^2}\rd k' = {1 \over 2 \pi}. 
\ee
The density of the gas is given by
\be
n= 2\int_{-K}^K \phi(k) \rd k, 
\ee
 while the ground state energy density is 
 \be
 E =-{c^2 n \over 4}+2  \int_{-K}^K k^2 \phi(k) \rd k. 
 \ee
 We will now introduce a convenient normalization due to \cite{tw1, tw2}. We introduce the function
 \be
 f(x)= 2 \pi \phi(c x), 
 \ee
 so that Gaudin's integral equation reads
\be
\label{gaudin-int}
{f(x) \over 2} +{1\over 2 \pi} \int_{-B}^B {f(y) \rd y  \over (x-y)^2+1} =1, \qquad -B<x<B. 
\ee
Here, $B=K/c$. The coupling (\ref{gamcn}) is given by
\be
\label{gB}
{1\over \gamma}={1\over \pi} \int_{-B}^B f(x) \rd x, 
\ee
while the normalized ground state energy density (\ref{egamma}) is
\be
\label{egam-ex}
e(\gamma)= -{\gamma^2\over 4}+ \pi^2 {\int_{-B}^B f(x) x^2 \rd x \over \left(\int_{-B}^B f(x)  \rd x\right)^3}. 
\ee
Let us note that, in the Bethe ansatz solution, the basic parameter is $B$, 
the endpoint of the interval where $f(x)$ is supported. The coupling constant $\gamma$ is a non-trivial function
of $B$, and $e(\gamma)$ depends on $\gamma$ through $B$. In addition, the weak coupling limit $\gamma \rightarrow 0$ correspond to the limit of large $B$. 

Unfortunately, the Gaudin integral equation can not be solved analytically. Therefore, although the Bethe ansatz solution makes it possible 
to calculate the ground state energy density exactly, one usually has to resort to numeric methods or to an asymptotic analysis in particular limits. It turns out that 
the study of the Gaudin equation in the weakly coupled regime is particularly difficult, since the integral equation becomes singular when $c\rightarrow 0$. The strong coupling 
regime is much easier to analyze, and one finds that as $\gamma \rightarrow \infty$, 
\be
\label{gamma-strong}
e(\gamma) \approx -{\gamma^2 \over 4} +{\pi^2 \over 48}+ \CO(\gamma^{-1}). 
\ee

The Bethe ansatz solution presented above describes the ground state with zero polarization, but it can be extended 
to arbitrary polarization. This makes it possible to calculate exactly the energy gap (also known as spin gap). The energy gap is 
defined as (minus) the energy difference between the ground state, in which there are equal numbers 
of spin up and spin down particles, and the first excited state, in which one breaks a Cooper pair 
and obtains in this way a triplet state with total spin $S=1$. It is easy to see from the Bethe ansatz solution that the spin gap 
can be calculated by solving an integral equation (see e. g. \cite{ko,zhou-exact}), and one finds, at leading order in a $\gamma$ expansion \cite{ko,frz}, 
\be
\label{spin-gap}
{\Delta \over E_F} \approx {16 \over \pi} {\sqrt{\gamma \over \pi}} \re^{-{\pi^2 \over 2 \gamma}}. 
\ee
Note that the exponent is identical to the one obtained in the BCS approximation to the gap (\ref{bcs-gap}). 

The case of a repulsive interaction with $c<0$ can be also solved with the Bethe ansatz, but one obtains 
a different integral equation for the ground state energy density (see \cite{sb-book} for a useful presentation). This equation can be put in the form 
\be
{f(x) \over 2} - \int_{-B}^B \CK(x-x') f(x') \rd x'= 1.  
\ee
The kernel appearing in this equation is
\be
\label{r-kernel}
\CK(x)= \int_\IR {\re^{-\ri p x} \over 1+ \re^{|p|}} {\rd p \over 4 \pi}. 
\ee
More explicitly, one has
\be
\CK(x)= {1\over 8 \pi } \left\{ \psi\left( 1+ {\ri x \over 2}\right)+ \psi\left( 1- {\ri x \over 2}\right)- \psi\left( {1\over 2} + {\ri x \over 2}\right)- \psi\left(  {1\over 2}-{\ri x \over 2}\right) 
\right\}, 
\ee
where $\psi(z)$ is the digamma function. Note that this kernel is no longer algebraic, and it is similar to the ones appearing in the Bethe equation for 
integrable field theories in two dimensions (like the $O(n)$ sigma model or the Gross--Neveu model). The coupling constant in the repulsive case is defined as 
\be
\gamma ={|c| \over n}
\ee
and it is given by
\be
\label{gammaB-rep}
{1\over \gamma}= {1\over 4 \pi} \int_{-B}^B f(x) \rd x. 
\ee
The ground state energy density is 
\be
e_r(\gamma)= {E \over n^3}= 16 \pi^2 { \int_{-B}^B  x^2 f(x) \rd x \over \left( \int_{-B}^B f(x) \rd x \right)^3}. 
\ee
As in the attractive case, it is difficult to use the integral equation above to study the weakly coupled gas. 
However, one can easily study the strongly coupled limit, and one finds (see e.g. \cite{guan-review} and references therein)
\be
\label{rep-sc}
e_r(\gamma) \approx {\pi^2 \over 3}, \qquad \gamma \rightarrow \infty. 
\ee

Finally, the Bethe ansatz solution for the ground state energy density can be extended to fermions with $\kappa$ spin components and an 
attractive interaction \cite{taka} (see e.g. \cite{guan-multi} for additional 
information and references). The density of Bethe roots satisfies now the integral equation
\be
\label{k-gaudin}
{f_\kappa(x) \over 2} + \int_{-B}^B \CK_\kappa(x-x') f_\kappa(x') \rd x'= {\kappa \over 2}, 
\ee
where the kernel is given by 
\be
\label{k-kernel}
\CK_\kappa(x)={1\over 2\pi}  \sum_{s=1}^{\kappa-1} {2 s/\kappa  \over (2 s /\kappa)^2 +x^2}.
\ee
The density $n$ and energy $E$ are now given by 
\be
{n \over \kappa}= {r\over 2 \pi}  \int_{-B}^B f_\kappa(x) \rd x, \qquad {E\over \kappa}=-{\kappa^2-1 \over 12 \kappa} c^2 n + {r^3  \over 2 \pi} \int_{-B}^B x^2 f_\kappa(x) \rd x,  
\ee
where
\be
r={c\kappa \over 2}.
\ee
The 't Hooft-like parameter $\lambda$ introduced in (\ref{lam-thooft}) is given by 
\be
{1\over \lambda}=  {1\over  \pi}  \int_{-B}^B f_\kappa(x) \rd x, 
\ee
and the rescaled energy density defined in (\ref{elambda}) can be obtained from 
\be
\label{elam-ex}
e(\lambda; \kappa)= -{\kappa^2 -1 \over 3 \kappa^2} \lambda^2+ \pi^2 {\int_{-B}^B f_\kappa (x) x^2 \rd x \over \left(\int_{-B}^B f_\kappa(x)  \rd x\right)^3}.
\ee

\begin{figure}[tb]
\begin{center}
\begin{tabular}{cc}
\resizebox{60mm}{!}{\includegraphics{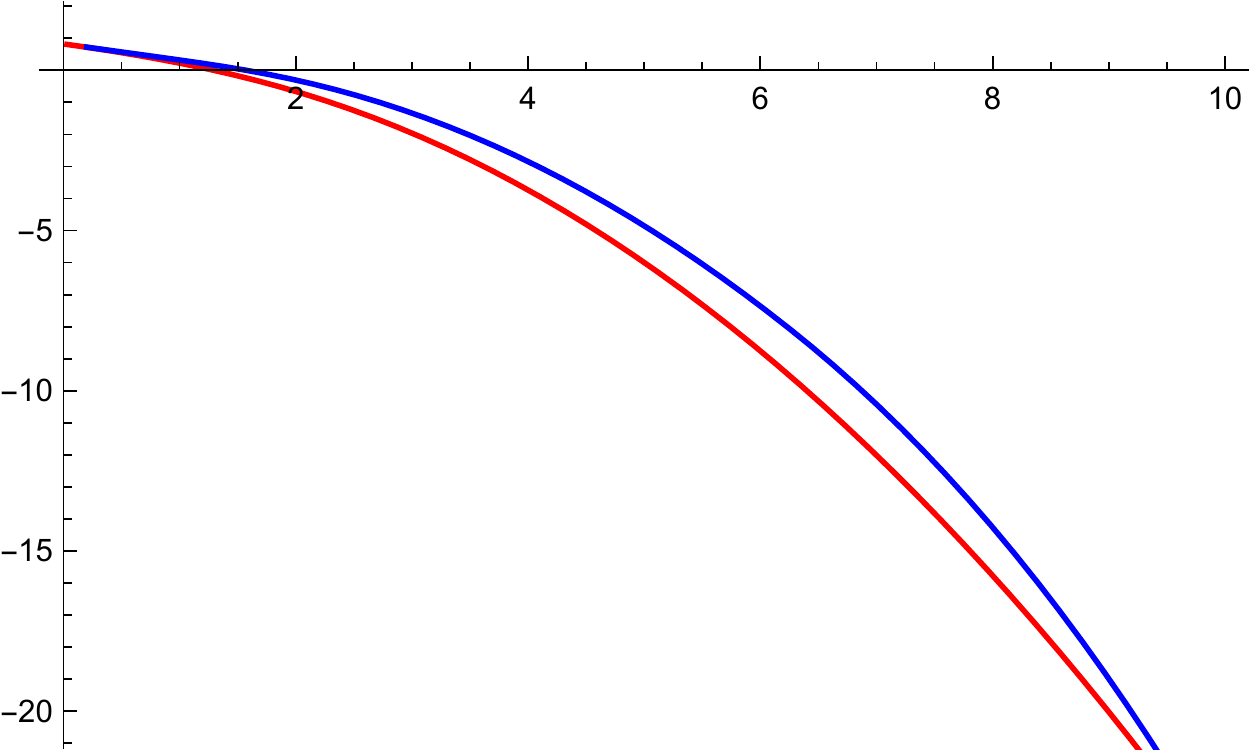}}
\hspace{3mm}
&\qquad \qquad 
\resizebox{60mm}{!}{\includegraphics{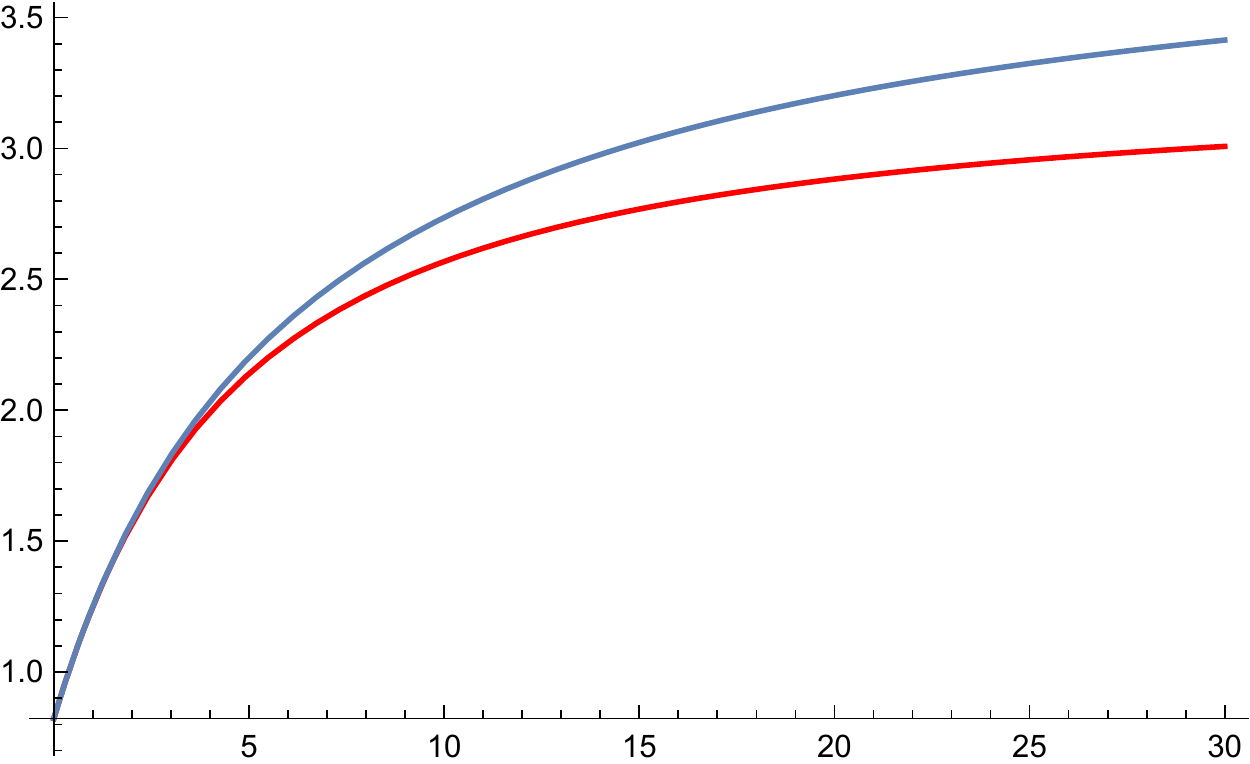}}
\end{tabular}
\end{center}
  \caption{Comparing the exact Bethe ansatz solution of the Gaudin--Yang model to approximation schemes. In the figure on the left, we consider the attractive case and we compare the exact answer for the 
  ground state energy density $e(\gamma)$ (bottom, in red), with the BCS mean field approximation (top, in blue), as a function of $\gamma$. In the figure on the right, we consider the repulsive case and we compare $e_r(\gamma)$ (bottom, in red) with the pp ladder resummation (top, in blue). 
}
\label{comps}
\end{figure}

One can use the Bethe ansatz exact solution to assess the accuracy of well-known approximations to the calculation of the ground state energy density. For example, in \cite{montse,quick}, the 
exact solution was compared to the BCS result in the attractive case. The BCS approximation provides a powerful weak/strong coupling interpolation procedure, and one can see that 
it is very successful, as illustrated in \figref{comps}. As noted in \cite{montse, quick}, it matches exactly both the weak coupling and the strong coupling limits. 

The resummation of diagrams also provides a mechanism to interpolate between weak and strong coupling. In \cite{schafer,kaiser1}, the resummation of pp ladder diagrams in the three-dimensional 
Fermi gas with a contact interaction was studied, and it was used to find an estimate of the Bertsch parameter of the unitary gas. A similar calculation can be done in the 
repulsive Gaudin--Yang model. The ladder resummation (\ref{e-pp-ladder}) provides an estimate for the ground state energy, $e_{\rm pp}(-\gamma)$, which can be compared to the exact 
answer $e_r(\gamma)$ obtained from the Bethe ansatz solution. This is illustrated in \figref{comps}. In this respect, it is interesting to define the analogue of the Bertsch parameter $\xi$ for the repulsive 
Gaudin--Yang model, as 
\be
\lim_{\gamma \rightarrow \infty} e_r(\gamma) = \xi e_0. 
\ee
In view of (\ref{rep-sc}), the Bethe ansatz gives the exact answer
\be
\xi_{\rm exact}= 4, 
\ee
while the pp ladder approximation gives
\be
\xi_{\rm pp}=1-  \int_0^1 \rd s \int_0^{1-s} \rd t {\pi^2 \over F_{\rm pp}(s,t)}= 4.84...
\ee
On the other hand, this approximation is not suitable for the attractive regime, and it fails to capture even qualitatively the strong coupling behavior (\ref{gamma-strong}). 

\section{Exact results for the perturbative series}

\label{sec-pert-series}

As we mentioned in the Introduction, the analysis of the Gaudin integral equation at weak coupling turns out to be highly non-trivial, 
since in this limit the integral equation becomes singular. A similar problem occurs 
in the Lieb--Liniger model, as it was already noted in \cite{ll}. As a result, it is very difficult to determine analytically the perturbative series for the ground state 
energy density of these models. In spite of numerous efforts \cite{iw-gy,tw1,tw2}, 
the series has only been derived up to second order in the coupling constant (even at the physical level of rigor). More 
recently, the combination of numerical analysis and inspired guesswork has led to conjectures for 
the exact form of some of the additional coefficients \cite{prolhac,lang}. 

In this section, we will solve this longstanding problem and present a method that determines systematically the 
perturbative series for the ground state energy density of the Gaudin--Yang model from a direct analysis of the integral equation\footnote{This method 
can be also used to obtain the perturbative series for the Lieb--Liniger model, which is presented in a separate paper \cite{mr-ll}.}. The method goes back 
to the work of Popov on the Lieb--Liniger equation \cite{popov} (and even earlier, to the work by Hutson on the circular plate condenser \cite{hutson}), but it was substantially improved by D. Volin in a 
remarkable {\it tour de force} \cite{volin,volin-thesis}. Volin's work focused on Bethe ansatz equations 
appearing in the context of integrable field theories in two dimensions (in particular the $O(n)$ sigma model). Gaudin's integral equation is an even simpler 
example where this method can be applied.

\subsection{Resolvent in the bulk and edge regimes}

\label{resolvent-section}

The method we will use has three fundamental ingredients. The first one is to rewrite the integral equation as a difference equation for the 
{\it resolvent} of the density of Bethe roots. The second one is to consider two different regimes for the equation: the bulk regime near the origin, and the edge regime near the 
endpoints of the support. One then matches asymptotic expansions in the two regimes (this was already the main idea in \cite{hutson,popov}). The third ingredient 
involves finding an explicit solution for the difference equation in the edge regime, after a Laplace transform. 

We will now apply the method to Gaudin's integral equation. As we will see in the next section, it is also possible 
to appeal to the conventional Wiener--Hopf method to obtain the wished-for solution for the resolvent in the edge regime without even writing the difference equation. This makes it possible 
to analyze a more general family of integral equations. 

Let us first write the kernel of Gaudin's integral equation (\ref{gaudin-int}) as 
\be
\label{ko}
{1\over x^2+1}= \mO {1\over x}, 
\ee
where
\be
\mO= -{1\over 2 \ri}\left( \mD - \mD^{-1}\right)
\ee
and 
\be
\mD= \re^{\ri \partial_x}
\ee
is the shift operator which acts as $x \rightarrow x+ \ri$. Following \cite{volin,volin-thesis} (see also \cite{ksv}), we introduce the resolvent of the density of Bethe roots as 
 \be
 R(x)= \int_{-B}^B {f(x') \over x-x'} \rd x'. 
 \ee
 This function is analytic in the complex $x$-plane except for a discontinuity in the interval $[-B, B]$, given by 
 \be
 f(x)=-{1\over 2 \pi \ri} \left( R(x+ \ri \epsilon)-R(x-\ri \epsilon) \right). 
 \ee
 Let us note that $R(x)$ is an {\it odd} function on the real axis away from $[-B, B]$, while its discontinuity is an {\it even} function of $x$. From its definition we deduce that 
 \be
 R(x)= \sum_{k \ge 0} \langle x^k\rangle  x^{-k-1} , \qquad\langle x^k\rangle= \int_{-B}^B f(x) x^k \rd x. 
 \ee
In particular, by using (\ref{gB}) we can compute the coupling constant $\gamma$ as a function of $B$ from the residue at infinity of the resolvent. 
It is now easy to see that Gaudin's integral equation leads to the following difference equation for $R(x)$:
\be
\label{gy-eq}
(1+\mD) R(x+ \ri \epsilon)- (1+ \mD^{-1}) R(x-\ri \epsilon) =-4 \pi \ri. 
\ee

The weak coupling limit of the Gaudin--Yang model corresponds to the limit of large $B$. Therefore, a natural strategy is to study the resolvent 
in a systematic expansion in $1/B$. To do this, we study the resolvent in two different regimes. The first one is the so-called {\it bulk regime}, in which we take the limit
 \be
 \label{Btheta}
 B\rightarrow \infty, \qquad x \rightarrow \infty,
 \ee
 in such a way that 
 \be
 u={x \over B}
 \ee
 is fixed. This is therefore appropriate to study $f(x)$ near $x=0$. The second regime is the so-called {\it edge regime}, in which we also have (\ref{Btheta}) but we keep fixed the variable
 \be
 \label{z-var}
 z=2\left(x-B\right). 
 \ee
This is therefore appropriate to study $f(x)$ near the edge of the distribution $x=B$. 
 
 For the bulk regime, and inspired by \cite{hutson,popov,iw-gy,volin}, we propose the following ansatz:
\be
\label{solgy}
R(x)= -\log\left( {x-\!B \over x+\!B} \right) +\sum_{n=1}^\infty \sum_{m=0}^\infty \sum_{k=0}^{n+m} {c_{n,m,k}(x/B)^{p(k)} \over B^{m-n}\left(x^2-B^2\right)^{n}} \, \left(\log\left( {x\!-\!B \over x\!+\!B} \right) \right)^k, 
 \ee
where $p(k)$ is $0$ or $1$ and 
\be
p(k)=k-1 \qquad  \text{mod $2$}.
\ee
Further justification for this ansatz will be clear when we study the edge regime. Note that $R(x)$ is indeed an odd function on $\IR\backslash [-B, B]$. 
Its discontinuity at the interval $[-B, B]$ leads to the following ansatz for $f(x)$, 
\be
f(x)=1 - \sum_{n=1}^\infty \sum_{m=0}^\infty \sum_{k=0}^{n+m} {(-1)^n c_{n,m,k}(x/B)^{p(k)} \over B^{m-n}\left(B^2-x^2\right)^{n}} \sum_{r=0}^{[{k-1 \over 2}]} {k \choose 2r+1} (-1)^r \pi^{2r} \left(\ell(x) \right)^{k-2r-1}.
\ee
where 
 \be
 \ell(x)= \log \left|  {x-B \over x+B} \right|, 
 \ee
 and $x \in [-B, B]$. The coefficients $c_{n,m,k}$ have all the information needed to compute the ground state energy. First of all, one has that
\be
\label{rho-exp}
{1\over \gamma}={2 B \over \pi}+{1\over  \pi} \sum_{m \ge 0} {c_{1,m,0} \over B^m}. 
\ee
We also have, 
\be
\label{x2-exp}
\langle x^2 \rangle= {2 B^3 \over 3}+ B^2 \sum_{m \ge 0} {c_{1,m,0}- 2 c_{1,m,1} +4 c_{1,m,2} \over B^m} + B \sum_{m \ge 0} { c_{2,m,0} \over B^m}. 
\ee
The coefficients $c_{n,m,k}$ can be partially determined by using the difference equation (\ref{gy-eq}), but this is not enough to fix their value. We then look at the edge regime. 

To study the edge regime, we write $R(z)=R(x(z))$ as a Laplace transform, 
\be
\label{laplace}
R(z) = \int_0^\infty \hat R(s) \re^{-s z} \rd s. 
\ee
As we will see in a moment, $R(s)$ decreases as $1/s$ at infinity, and the Bromwich inversion formula reads
\be
\label{inv-laplace}
\hat R(s)= \int_{-\ri \infty+\epsilon}^{\ri \infty+ \epsilon} \re^{s z} R(z) {\rd z \over 2 \pi \ri}. 
\ee
 The inverse Laplace transform of (\ref{gy-eq}) is obtained by replacing 
\be
\mD^a= \re^{2 \pi \ri a \partial_z} \rightarrow \re^{-2 \pi \ri a s}, 
\ee
and it reads
\be
\label{laplaceequation}
\cos\left(s\right) \left( \re^{-\ri s} \hat R(s-\ri \epsilon)-\re^{\ri s} \hat R(s+\ri \epsilon) \right)={1\over s-\ri \epsilon} -{1\over s+\ri \epsilon}, \qquad s<0.  
\ee
To find the correct solution to (\ref{laplaceequation}) we demand the following properties for $\hat R(s)$ at each order in a $1/B$ expansion:

 \begin{enumerate}
 
 \item $\hat R(s)$ is analytic everywhere except on the negative real axis.
 
 \item $\hat R(s)$ has simple poles at $s/\pi=-n-1/2$, where $n$ is a non-negative integer. 
 
 \item $\hat R(s)$ has an expansion as $s\rightarrow \infty$ in integer, negative powers of $s$. 
 
 \end{enumerate}
 These analyticity properties can be justified as in \cite{volin, volin-thesis}. 
 Note that the structure of zeroes in the resolvent follows from (\ref{laplaceequation}). 
 The third property follows from the following consideration. The function $f(x)$ is regular at $x=B$, and therefore it has an expansion 
\be\label{taylor}
f(x)=f_0+(x-B) f_1+(x-B)^2 f_2+\cdots .
\ee
To reproduce this structure through the discontinuity formula we must have
\be
\label{taylor2}
R(z)=-\log(z)\left(f_0+{ z \over 2} f_1+{z^2 \over 4} f_2+\cdots \right). 
\ee
We now use the following (analytically continued) Laplace transform, 
\be
s^{-n}  \leftrightarrow {(-1)^n \over (n-1)!} \log z \, z^{n-1}. 
\ee
In this way, the inverse Laplace transform of (\ref{taylor2}) gives
\be
\label{lt-taylor2}
\hat R(s)={f_0 \over s}-{f_1 \over 2 s^2}+{f_2 \over 4 s^3}+\cdots,
\ee
which is precisely the third property above. The connection between (\ref{taylor}) and (\ref{lt-taylor2}) is very useful since it makes it possible to test 
the solution for $\hat R(s)$ against a numerical study of the distribution $f(x)$ at the edge. 

The most general solution of (\ref{laplaceequation}) which satisfies the analyticity properties stated above is the following:
 \be
 \label{sol1}
  \hat R(s)=\Phi(s)\left(\frac{1}{s}+Q(s)\right),
  \ee
  where
\be
\label{exact-phi}
\Phi(s)={1\over {\sqrt{\pi}}} \exp \left[ {s \over  \pi} \log\left( { \pi \re \over s} \right) \right] \Gamma\left( {s\over  \pi}+{1\over 2} \right),
\ee
and $Q(s)$ is given by
  \be
  Q(s)=\frac 1{Bs} \sum_{n,m=0}^\infty { Q_{n,m}(\log B) \over B^{m+n} s^{n}}.
\ee
 The dependence of $Q(s)$ on $B$ is a consequence of the structure of (\ref{solgy}). We note that the coefficients $Q_{n,m}$ are undetermined functions of $\log B$. 
 The overall coefficient in $\Phi(s)$ is fixed by studying (\ref{laplaceequation}) near $s=0$. In the next section we will give a different argument, based on the Wiener--Hopf method, 
 to obtain the function $\Phi(s)$. 
  
It turns out that the information contained in (\ref{exact-phi}) is enough to compute the coefficients $c_{n,m,k}$ and $Q_{n,m}$, therefore the ground state energy density of the Gaudin--Yang model. This is done 
by ``matching" the bulk and the edge regimes, as already proposed in \cite{hutson, popov}. In Volin's reformulation of the method, we first consider the ansatz (\ref{solgy}) for $R(x)$ in 
the bulk regime, by setting $x=z/2 +B$ and expanding around $B=\infty$, $z/B =0$. We do the Laplace transform of the resulting expression, and we compare the result to the expansion of (\ref{sol1}) around $s=0$. 
This turns out to fix all the coefficients. In Appendix \ref{match} we work out explicitly the very first terms to illustrate the procedure. Note that the terms obtained in the expansion of (\ref{sol1}) provide a further 
check of the bulk ansatz (\ref{solgy}). The general expressions (\ref{rho-exp}) and (\ref{x2-exp}) give, after using the values for the coefficients obtained in (\ref{res-coefs}), 
\be
\ba
\label{rhox-exp}
{1\over \gamma}&={2 B \over \pi}+ {1\over  \pi^2} \left( \log(\pi B) +1 \right) +{1\over 2 \pi^3 B} \left( \log(\pi B) + {1 \over 2} \right) +\CO(B^{-2}), \\
\langle x^2 \rangle &= {2 B^3 \over 3}+{B^2\over \pi} \left( \log(\pi B)-1\right)+ {B \over 2 \pi^2} \left( \log^2(\pi B)- \log(\pi B)-{5\over 2} +{2 \pi^2\over 3}\right) +
\CO(B^0). 
\ea
\ee
They are in agreement with the results in \cite{tw1, tw2}. We also obtain, for the distribution $f(x)$ in the bulk regime, 
\be
\ba
f(x)&= 1 +{1\over \pi B (1-u^2)}\\
& +{1 \over \pi^2 B^2} \left( {1 +  \log(\pi B)  \over 2  (1-u^2)} -{\log(B \pi) \over  (1-u^2)^2}- {u \log\left({1-u \over 1+u} \right) \over (1-u^2)^2}  \right)+ \CO\left(B^{-3} \right), \\
\ea
\ee
where $u =x/B$. The leading order term in this expression reproduces the result of \cite{iw-gy}. 

As in \cite{volin, volin-thesis}, the recursive procedure to determine the coefficients $c_{n,m,k}$ can be fully automatized in a symbolic program, and one can push the 
calculation to any desired order, the only limitation being CPU time. We obtain in this way two series in $1/B$, one for $\rho$ and another for $\langle x^2\rangle$, whose coefficients are polynomials in $\log B$. This leads to a series of the same type for $\gamma$, which can be inverted to obtain $1/B$ as a power series in $\gamma$ (involving as well $\log \, \gamma$). One can then re-express the r.h.s. of (\ref{egam-ex}) as a power 
series in $\gamma$ in order to calculate the perturbative series for the ground state energy density. It turns out that the dependence on $\log \, \gamma$ fully disappears in the 
final result (it was erroneously claimed in \cite{ko} that this is not the case). This is similar to what has been observed in integrable field theories in \cite{fnw1,bbbkp,volin, volin-thesis}.

In this way one finds, up to order $\gamma^{10}$, 
\be
\label{eten}
\ba
& e(\gamma)=\frac{\pi ^2}{12}-\frac{\gamma }{2}-\frac{\gamma ^2}{12}-\frac{ \zeta (3)}{\pi ^4}\gamma ^3-\frac{3  \zeta (3)}{2 \pi ^6}\gamma ^4-\frac{3 \zeta (3)}{\pi ^8}\gamma ^5 -\frac{5  (5 \zeta (3)+3 \zeta (5))}{4 \pi ^{10}}\gamma ^6\\&-\frac{3\left(12 \zeta (3)^2+35 \zeta (3)+75 \zeta (5)\right)}{8 \pi ^{12}} \gamma ^7 -\frac{63  \left(12 \zeta (3)^2+7 \zeta (3)+35 \zeta (5)+12 \zeta (7)\right)}{16 \pi ^{14}}\gamma ^8\\&-\frac{3  \left(404 \zeta (3)^2+240 \zeta (5) \zeta (3)+77 \zeta (3)+735 \zeta (5)+882 \zeta (7)\right)}{4 \pi ^{16}}\gamma ^9\\
&-\frac{27  \left(160 \zeta (3)^3+1800 \zeta (3)^2+3720 \zeta (5) \zeta (3)+143 \zeta (3)+2310 \zeta (5)+6363 \zeta (7)+1700 \zeta (9)\right)}{32 \pi ^{18}}\gamma ^{10}\\&+{\cal O}\left(\gamma ^{11}\right)   \ea
   \ee
Note that, up to $\gamma^3$, this agrees with the perturbative results obtained above. 
The coefficients in (\ref{eten}) are in accord with the numerical calculations of 
S. Prolhac \cite{prolhac}. We have pushed the calculation up to order $50$ in $\gamma$, which makes it possible to 
determine the large order behavior of the series with some precision.

\subsection{Relation to the Wiener--Hopf method}

One popular approach to study the integral equations appearing in the Bethe ansatz is to use the Wiener--Hopf method (see e.g. \cite{hmn,hn,fnw1} for examples in the context 
of relativistic field theories). In the cases we are interested in here, the integral equation
is of the form
\begin{equation}
f(x)-\int_{-B}^B \CK (x-x')f(x') \rd x'= \alpha, 
\label{origTBA}
\end{equation}
where $\alpha$ is a real constant. In the Wiener--Hopf method, one extends this equation to the whole real axis in order to use 
Fourier transform techniques. This method seems however to be less efficient than the one presented here. For example, in the study of the 
Gaudin--Yang model in \cite{tw1,tw2} with the Wiener--Hopf method, only the terms up to order $\gamma^2$ are obtained, in spite of some effort. 
A key aspect of our approach is the explicit solution (\ref{exact-phi}), since this provides the initial data 
that allow to fix all the constants. This function is a solution to the discontinuity equation satisfied by $\hat R(s)$. 
However, as noted already in \cite{volin}, one can see in many examples 
that $\Phi(s)$ is closely related to the Wiener--Hopf decomposition of the kernel $\CK(x)$. This makes it possible to obtain $\Phi(s)$ 
without the need to determine and solve the equations satisfied by the resolvent. We will now give a heuristic argument to establish this relation. 
The argument below is only valid when $G_\pm (0)$ is finite and provides information about the strict large $B$ limit, but it can be extended to more 
general situations. 

\begin{figure}
\center
\includegraphics[height=5cm]{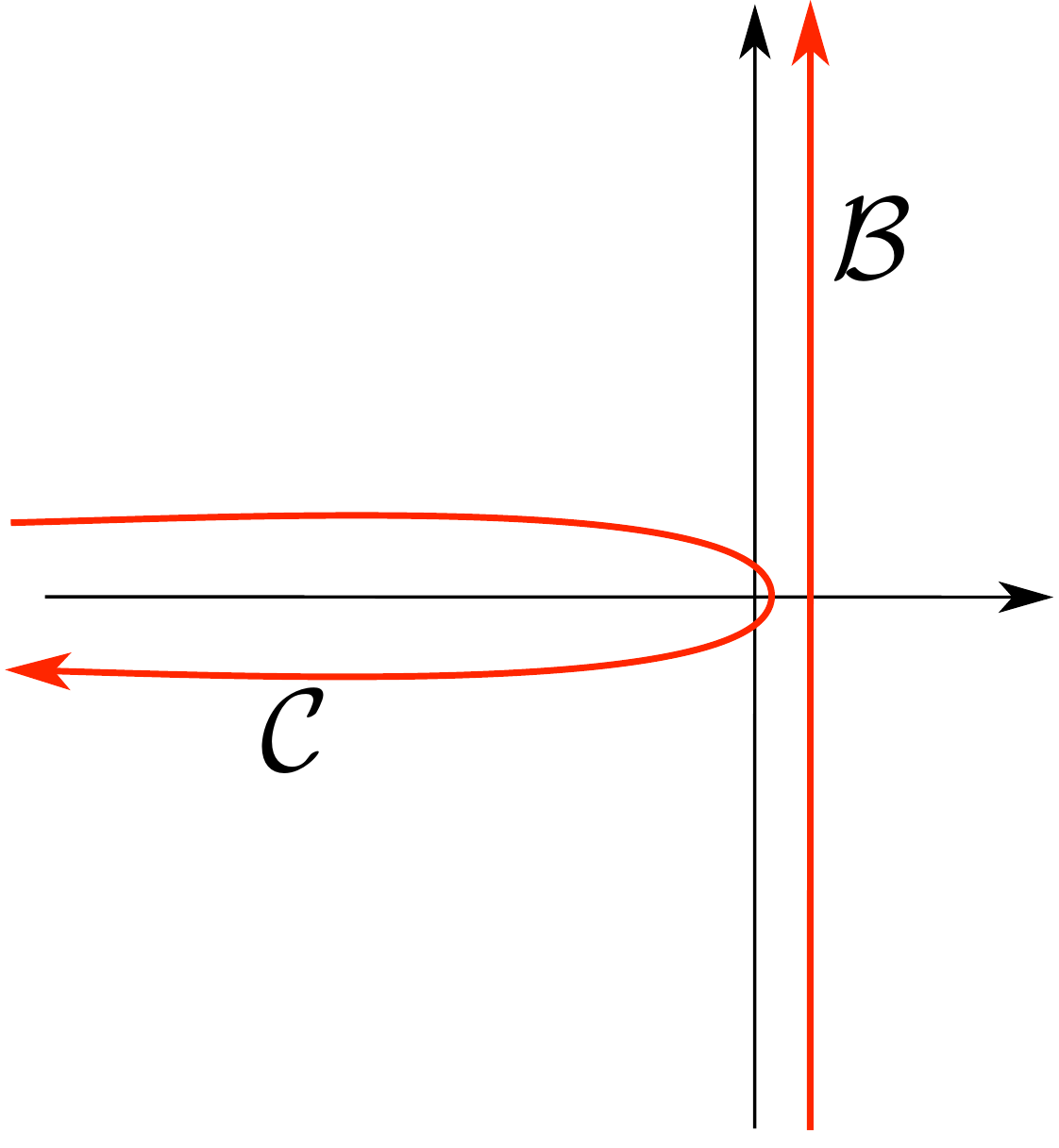}
\caption{The Bromwich contour $\CB$ appearing in the inverse Laplace transform (\ref{inv-laplace}) can be deformed into (minus) the contour $\CC$.}
\label{c-contour}
\end{figure}

Let us write the integral equation (\ref{origTBA}) in the edge regime, in terms of the variable (\ref{z-var}), as
\begin{equation}
\label{edge-ie}
\int_{-\infty}^0\left[\delta(z-z')-K(z-z')\right]f(z')\rd z'= \alpha\,,\quad  z<0. 
\end{equation}
Here, $f(z)=f(x(z))$ and $K(z)=\CK(z/2)/2$, in order to absorb the multiplicative factor from the change of variables. 
By extending the above equation to all real $z$, we obtain
\be
\int_{-\infty}^\infty \left[\delta(z-z')-K(z-z')\right]f(z')(1-\theta(z'))\rd z '= \,\alpha+\theta(z)\xi(z), 
\addtocounter{equation}{1}\tag{\theequation} \label{eq_preF}
\ee
where $\xi(z)$ is {\it a priori} an unknown function. Let us define the following Fourier transforms
\be
\label{eq_fminus}
\ba
K(\omega)&=\int_{-\infty}^\infty \re^{\ri\omega z} K(z) \rd z\,,\\
F_-(\omega)&=\int_{-\infty}^\infty \re^{\ri\omega z} f(z)(1-\theta(z)) \rd z=\int_{-\infty}^0 \re^{\ri\omega z} f(z) \rd z, \\
X_+(\omega)&=\int_{-\infty}^\infty \re^{\ri\omega z} \xi(z)\theta(z) \rd z=\int^{\infty}_0 \re^{\ri\omega z} \xi(z) \rd z\,.
\ea
\ee
The subscripts $\pm$ denote that something is analytic in the upper/lower half complex plane (including the real axis but possibly excluding the origin). We also introduce 
\be
1-K(\omega)={1\over G_+(\omega)G_-(\omega)}. 
\ee
When $K(\omega)$ is an even function of $\omega$, we have,  
\be
G_+(\omega)=G_-(-\omega).
\ee
 This decomposition can almost always be done provided $1-K(\omega)$ is well defined on the real axis, though some care 
 might be necessary at $\omega=0$. The integral equation becomes, 
 \begin{equation}
\frac{F_-(\omega)}{G_+(\omega)G_-(\omega)} =  \alpha 2 \pi \delta(\omega) +  X_+(\omega)\,.
\label{eq_WH0}
\end{equation}
By using that
\be
 2 \pi \ri \delta(\omega)= {1\over \omega-\ri\epsilon}-{1\over \omega+\ri\epsilon},
 \ee
  we can rewrite \eqref{eq_WH0} as
\begin{align}
\frac{F_-(\omega)}{G_-(\omega)} -\frac{\alpha}{\ri} \frac{G_+(0)}{\omega-\ri\epsilon}= -\frac{\alpha}{\ri}{ G_+(0) \over \omega+\ri\epsilon}+G_+(\omega)X_+(\omega)= C(\omega)\,.
\label{eq_WH}
\end{align}
From \eqref{eq_WH} it follows that $C(\omega)$ must be analytic in both the upper and the lower half complex planes, reducing it to an entire function. 
However, in their respective half planes (including the real axis), $F_-(\infty)=0$, $X_\pm(\infty)=0$ and $G_\pm(\infty)$ is a constant 
(the first two results follow from their definitions and (\ref{eq_preF}), while the latter can be checked explicitly in examples). 
Both sides are thus bound at infinity, and by Liouville's theorem, we conclude that $C(\omega)=0$. Thus, in the strict limit $B \rightarrow \infty$, we have
\begin{equation}
F_-(\omega)= \alpha G_+(0)\, \frac{G_-(\omega)}{\ri \, \omega}\,.
\label{eq_WH_basic}
\end{equation}
We can now relate $F_-(\omega)$ to $\hat{R}(s)$ introduced in (\ref{laplace}) (again in the strict large $B$ limit). 
We consider $\hat{R}(\ri s)$ for ${\rm Im}(s)<0$ and deform the Bromwich contour $\CB$ in (\ref{inv-laplace}) around the negative real axis, without crossing it. We obtain in this way minus the 
contour ${\cal C}$ shown in \figref{c-contour}. Therefore, 
\be
\ba
\hat{R}(\ri s)&=\int_\CB \frac{\re^{\ri s z}  }{2\pi \ri}R(z) \rd z=-\int_{\cal C} \frac{\re^{\ri s z}  }{2\pi \ri}R(z) \rd z\\
&=-\int_{-\infty}^{0}\re^{\ri s z} \left(\frac{R(z+\ri\epsilon)-R(z-\ri\epsilon)}{2\pi \ri}\right)\rd z=\int_{-\infty}^{0}\re^{\ri s z} f(z) \rd z \\
&= F_-(s).
\ea
\ee
We conclude that
\be
\label{hatr-short}
\hat R(s) =\alpha G_+(0)\, \frac{G_-(-\ri s )}{s}, 
\ee
which is precisely of the form of the first term in the r.h.s. of (\ref{sol1}). In other words, 
\be
\label{phi-g}
\Phi(s) =\alpha G_+(0)\,G_-(-\ri s ).
\ee
The results (\ref{hatr-short}) and (\ref{phi-g}) show that we can skip the steps of finding and solving an equation for the resolvent. We 
can simply obtain the crucial leading term of $\hat R(s)$ from the Wiener--Hopf decomposition of the kernel, which is available in many situations. Note however 
that (\ref{eq_WH_basic}) gives the solution in the strict large $B$ limit. There are $1/B$ corrections to this result which lead to the term $Q(s)$ in (\ref{sol1}). 

As an example, let us reconsider Gaudin's integral equation. In the notation of (\ref{origTBA}) we have
\be
K(\omega)=- \re^{-2|\omega|}\,,\quad  \alpha=2, 
\end{equation}
and one has (see e.g. \cite{tw1})
\be
G_+(\omega)=\frac{1}{\sqrt{2 \pi }}\Gamma \left(\frac{1}{2}-\frac{\ri \omega}{ \pi }\right) \exp \left(\frac{\ri \omega \left(\log \left(-\frac{ \ri \omega}{\pi }\right)-1\right)}{ \pi }\right). 
\ee
It is easy to check that (\ref{phi-g}) reproduces the result (\ref{exact-phi}) for $\Phi(s)/s$. 

\subsection{The repulsive case}

The Gaudin--Yang model with a repulsive interaction is a good example where one can obtain the wished-for result for the resolvent by using (\ref{hatr-short}): the kernel has a very simple 
expression after Fourier transformation, and one can also find easily its Wiener--Hopf decomposition. In the notations of the previous section, one has
\be
K(\omega)=\frac{1}{1+\re^{2|\omega|}}\,,\quad \alpha=2, 
\ee
and 
\be
 G_+(\omega)=\frac{\sqrt{2 \pi }}{\Gamma \left(\frac{1}{2}-\frac{\ri \omega}{ \pi }\right) }\exp \left(-\frac{\ri \omega \left(\log \left(-\frac{ \ri \omega}{\pi }\right)-1\right)}{ \pi }\right)\,.
\end{equation}
From these results and (\ref{phi-g}) we obtain the function $\Phi(s)$. 
The ansatz for the resolvent in the bulk is the same as in (\ref{solgy}), with the only difference that the leading logarithmic term has an additional factor of $4$:
\be
R(x)= - 4 \log\left( {x-B \over x+ B} \right)+ \cdots 
\ee
The ansatz for the edge is (\ref{sol1}), albeit with the new $\Phi(s)$. Following the procedure of section \ref{resolvent-section}, we find that the ground state energy 
density is given by the series (\ref{eten}) after we change $\gamma \rightarrow -\gamma$, as expected. It is also possible to write down a difference equation for the 
resolvent of this model, by using the representation of the Gamma function given in \cite{volin, volin-thesis}.

\subsection{The Gaudin--Yang model with $\kappa$ spin components}

The Gaudin--Yang model with $\kappa$ spin components is particularly interesting, since we 
can consider the limit of large $\kappa$ and make contact with the RPA solution at subleading order in $1/\kappa$. 

To analyze the linear integral equation (\ref{k-gaudin}), we use the ingredients provided by the Wiener--Hopf method. We have, in the notation of (\ref{edge-ie}), 
\begin{equation}
K(\omega)=\frac{1-\re^{-4|\omega|}}{1-\re^{-4|\omega|/\kappa}},  \qquad 
\alpha=\kappa, 
\ee
and
\be
G_+(\omega)={1\over \sqrt{\kappa}}
{\Gamma\left(1-\frac{2 \ri \omega}{\pi}\right) \over \Gamma\left(1-\frac{2 \ri \omega}{\pi \kappa}\right)}
\exp\left[ 
{2 \ri \omega \over \pi} \left(  \log\left(-\frac{2 \ri \omega}{\pi}\right)-1-{1\over \kappa} \left(\log\left(-\frac{2 \ri \omega}{\pi \kappa}\right)-1\right)
\right)\right].
\ee
We can now use (\ref{phi-g}) to obtain $\Phi(s)$. The ansatz for the bulk and for the edge are the same as for the Gaudin--Yang model, with the only difference that in (\ref{sol1}) we have to use the new 
$\Phi(s)$. By using the method in section \ref{resolvent-section}, we can now proceed to calculate the ground state energy, as 
a function of $\lambda$ and $\kappa$ (we recall that $\lambda$ was defined in (\ref{lam-thooft})). It is convenient to present the results in terms of the large $\kappa$ expansion in (\ref{large-kappa}). One finds, 
for the first two series $e_n(\lambda)$, 
\be
\label{kappa-results}
\ba
e_1(\lambda)&=\lambda-\frac{\lambda^2}{3}-\frac{4  \zeta (3)}{\pi ^4}\lambda^3-\frac{12  \zeta (3)}{\pi ^6}\lambda^4-\frac{48
   \zeta (3)}{\pi ^8} \lambda^5-\frac{40  (5 \zeta (3)+\zeta (5))}{\pi ^{10}}\lambda^6\\
   &-\frac{120
   (7 \zeta (3)+5 \zeta (5))}{\pi ^{12}} \lambda^7-\frac{168  (21 \zeta (3)+35 \zeta (5)+4
   \zeta (7))}{\pi ^{14}}\lambda^8\\
   &-\frac{1344  (11 \zeta (3)+35 \zeta (5)+14 \zeta (7))}{\pi
   ^{16}}\lambda^9+\CO\left(\lambda^{10}\right), \\
   e_2(\lambda)&={\lambda^2 \over 3} +\frac{4  \zeta (3)}{\pi ^4}\lambda^3+\frac{24 \zeta (3)}{\pi ^6}\lambda^4 +\frac{144 \zeta
   (3)}{\pi ^8}\lambda^5 +\frac{40  (20 \zeta (3)+3 \zeta (5))}{\pi ^{10}}\lambda^6\\
   &+\frac{24 
   \left(175 \zeta (3)-6 \zeta (3)^2+100 \zeta (5)\right)}{\pi ^{12}}\lambda^7+\frac{168 
   \left(126 \zeta (3)-18 \zeta (3)^2+175 \zeta (5)+16 \zeta (7)\right)}{\pi
   ^{14}}\lambda^8\\
   &+\CO\left(\lambda^9\right). 
    \ea
   \ee
   The result for $e_1(\lambda)$ gives explicit predictions for the contributions of the first ring diagrams. In particular, we obtain a Bethe ansatz prediction for 
   the values of the integrals (\ref{il-int}). One finds, for example, 
\be
\CI_2= {\pi^3 \over 6}, \quad \CI_3={3 \pi \over 4} \zeta(3), \quad \CI_4= {3 \pi \over 4} \zeta(3), \quad 
\CI_5={15 \pi \over 16} \zeta(3), \quad \CI_6= {15 \pi \over 64} \left( 5 \zeta(3) + \zeta(5) \right).
\ee
The first two values can be verified analytically, since $\CI_2$ is proportional to the two-bubble diagram evaluated in $c_2^{\rm pp}$ in (\ref{cpp-values}), 
while the value of $\CI_3$ follows from the calculation in Appendix \ref{3-diagrams}. The remaining equalities can be verified numerically (although the direct numerical evaluation of $\CI_\ell$ becomes more 
and more difficult as $\ell$ is increased.) A detailed analysis of ring diagrams for a delta-function interaction in two dimensions has been made in \cite{kaiser2}. Interestingly, their evaluation 
is simpler than in the one-dimensional case, and it does not involve odd values of the zeta function. In the next section we will also examine the behavior of the $\CI_\ell$ as $\ell$ becomes large.

\sectiono{Large order behavior, energy gap, and renormalons}
\label{lo-sec}
\subsection{Large order behavior of the perturbative series}

The results obtained in the previous section make it possible to study the large order behavior 
of the perturbative series with numerical methods. 
We will first consider the series for $e(\gamma)$ in the original case of spin $1/2$, for which we know the exact form of the first $50$ terms. 

It is very easy to see that the coefficients in (\ref{e-cos}) grow factorially, i.e. $c_k \sim k!$. Following standard practice in the asymptotic analysis of perturbative series (see e.g. \cite{mmbook}), 
we assume the following ansatz:
\be
\label{lob}
c_k \sim {1 \over 2 \pi} A^{-b-k}  \Gamma(k+b) \left[ \mu_0 + {\mu_1 A  \over k+b-1} + {\mu_2 A^2 \over (k+b-2)(k+b-1)}+\cdots\right], \qquad k \gg 1.
\ee
The parameters $A$, $b$, and $\mu_i$, $i \ge 0$, can be extracted systematically from the sequence $c_k$, by a numerical analysis of appropriate auxiliary sequences 
(see e.g. \cite{msw}). For example, the parameter $A$ can be extracted from the sequence
\be
\label{sk-def}
s_k={k c_k \over c_{k+1}}, 
\ee
whose asymptotics is 
\be
s_k = A + \CO\left({1\over k}\right), \qquad k \gg 1. 
\ee
The precision in the determination of these parameters depends of course on the number of coefficients of the series which are available. In order to accelerate the convergence of the auxiliary sequences, 
we will use Richardson transforms \cite{bender-book,msw}. Our analysis gives
\be
\label{action}
A=\pi^2,
\ee
with a precision of six decimal digits, once Richardson transforms are used. This is illustrated in the left graphic of \figref{asym-graphics}. In addition, we have 
\be
\label{bmu0}
b=-1, \qquad \mu_0= -2. 
\ee
In other words, we find the asymptotic behavior 
\be
\label{asymck}
c_k \sim -{1\over \pi}(\pi^2)^{-k+1} \Gamma(k-1), \qquad k \gg 1.  
\ee
Our numerical analysis also indicates that 
\be
\mu_1 \approx -{3 \over 2 \pi^2}, 
\ee
although this is somewhat less precise. In the graphic on the right in \figref{asym-graphics} we show the sequence 
\be
\label{rkaux}
r_k = -{\pi^{2k-1} c_k \over \Gamma(k-1)}, 
\ee
together with its fifth Richardson transform. According to (\ref{asymck}), this sequence should asymptote $1$, as we observe in the graphic on the right in \figref{asym-graphics}. 

\begin{figure}[tb]
\begin{center}
\begin{tabular}{cc}
\resizebox{60mm}{!}{\includegraphics{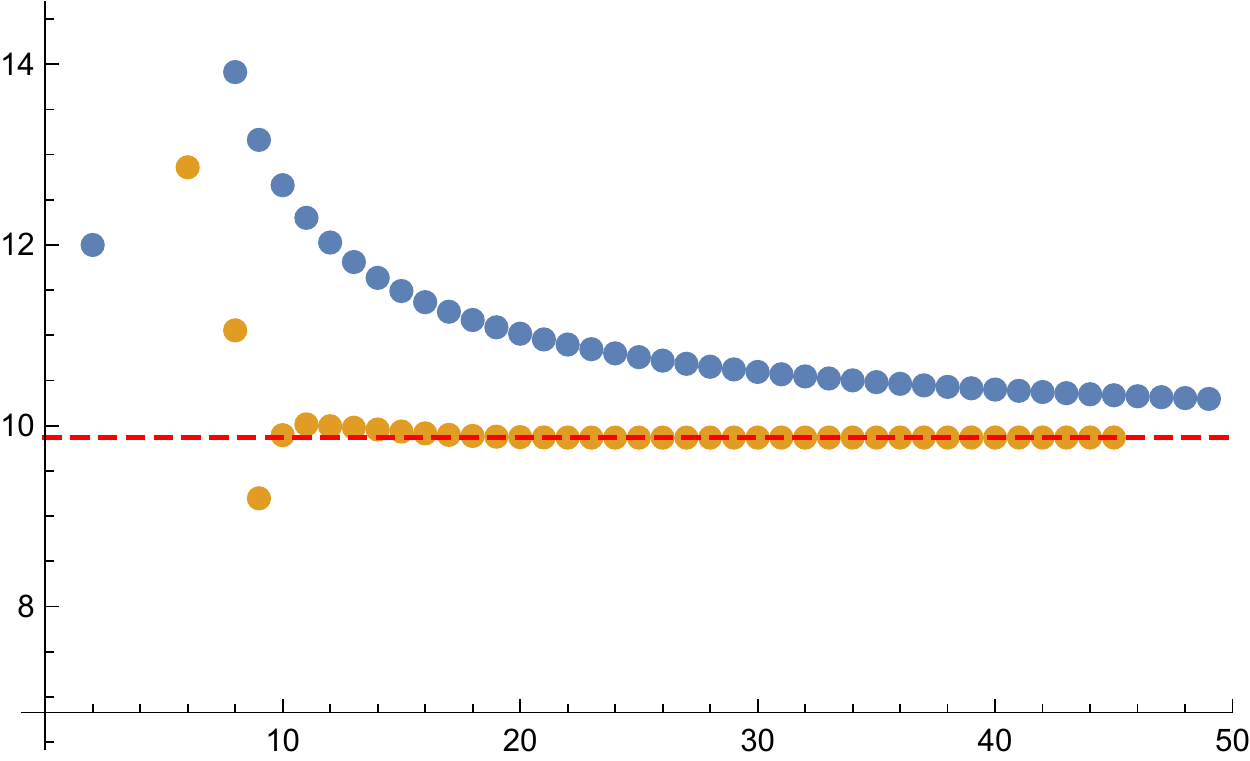}}
\hspace{4mm}
\qquad\qquad
&
\resizebox{60mm}{!}{\includegraphics{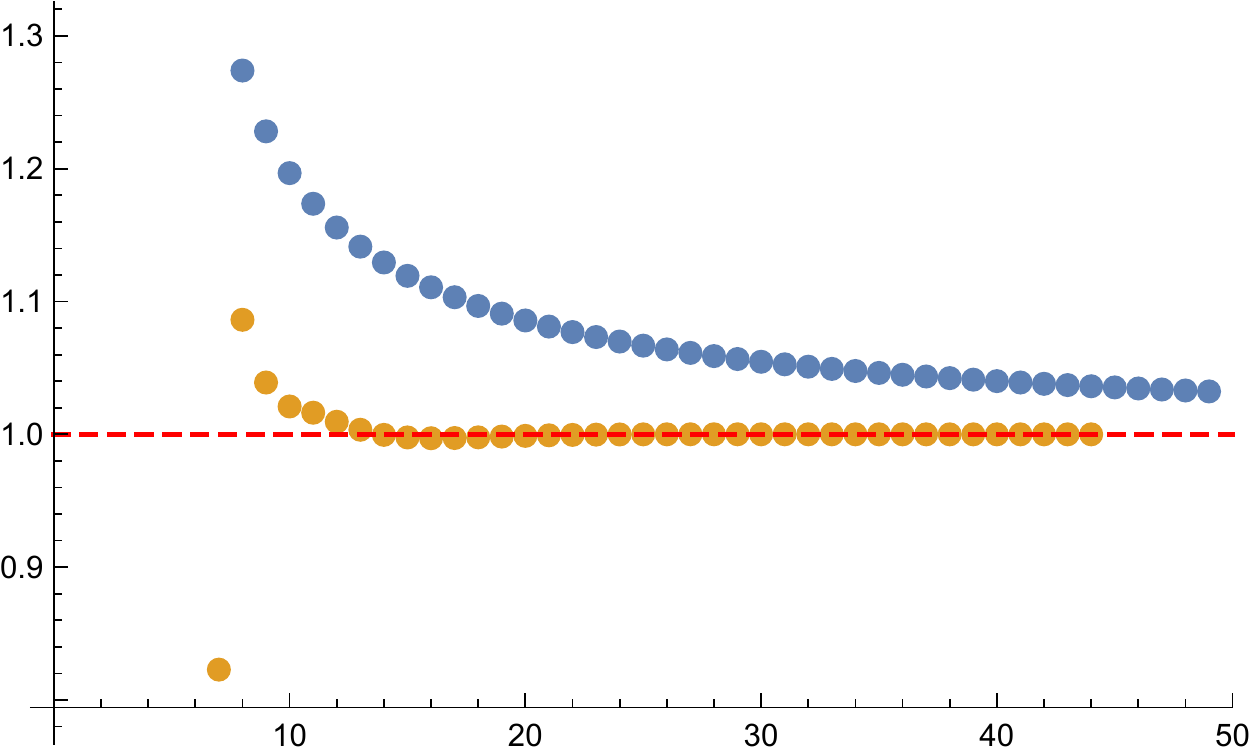}}
\end{tabular}
\end{center}
  \caption{In the figure on the left, we show the sequence $s_k$, defined in (\ref{sk-def}), while the sequence below is its fourth Richardson transform. The dashed line is the value $A=\pi^2$. The figure on the right 
  shows  the auxiliary sequence (\ref{rkaux}), which is the quotient of the $c_k$ by its leading asymptotics, together with its fifth Richardson transform below it. These sequences should approach the dashed line at $1$. 
}
\label{asym-graphics}
\end{figure}
%


These empirical results show that the perturbative series of the Gaudin--Yang model is factorially divergent. In the {\it attractive} case, the series is non-alternating, therefore 
not Borel summable. If we consider the Borel transform of $e(\gamma)$, 
\be
\widehat e(\zeta)= \sum_{k \ge 0} {c_k \over k!} \zeta^k, 
\ee
the large order behavior (\ref{asymck}) indicates that this function has its first Borel singularity at 
\be
\zeta= \pi^2. 
\ee
This is precisely twice the strength of the exponentially small term in the energy gap (\ref{spin-gap}), and 
it confirms our conjecture in the case of the Gaudin--Yang model. As we noted in the Introduction, 
the factor of two can be justified heuristically by the fact that the non-analytic correction to the energy in the BCS approximation is proportional to $\Delta_{\rm BCS}^2$, as we saw in (\ref{ebcs-ex}). 
 
Let us now recall some basic facts of the theory of Borel--\'Ecalle resummation (see e.g. \cite{mmbook}). 
When the Borel transform $\widehat \varphi(\zeta)$ of a perturbative series $\varphi(z)$ has singularities on 
the positive real axis, one can define lateral Borel resummations as 
\be
\label{lateralborel-theta}
s_{\pm} (\varphi)(z)=z^{-1} \int_{\CC_{\pm}} \rd \zeta \, \re^{-\zeta/z} \widehat \varphi (\zeta), 
\ee
where $\CC_\pm$ are integration paths slightly above (respectively, below) the positive real axis. If the coefficients of the perturbative series are real, 
the two lateral Borel resummations have the same real part but imaginary parts with opposite sign. On the other hand, the large order behavior (\ref{lob}) is associated to a 
non-perturbative ambiguity or non-perturbative correction of the form:
\be
\label{npa-gen}
\ri \, \re^{-A/z} z^{-b} \sum_{n=0}^\infty \mu_n z^n. 
\ee
This ambiguity gives the leading contribution to $s_{+} (\varphi)(z)-s_{-} (\varphi)(z)$. 

In the case of the Gauding--Yang model, it follows from (\ref{action}), (\ref{bmu0}) and (\ref{npa-gen}) that the non-perturbative ambiguity has the following form,  
\be
\label{np-amb}
-2 \gamma \, \re^{-\pi^2/\gamma}, 
\ee
at leading order in $\gamma$ and $\re^{-\pi^2/\gamma}$ as $\gamma \rightarrow 0$. 
This implies that the lateral Borel resummation of the perturbative series for $e(\gamma)$ has an 
imaginary piece given by 
\be
\label{im-ex}
{\rm Im}\, s_{\pm} (e)(\gamma) \sim \mp \gamma \, \re^{-\pi^2/\gamma}, \qquad \gamma \rightarrow 0. 
\ee

Let us consider a numerical example to illustrate this result. It is convenient to parametrize $\gamma$ by $B$, following (\ref{gB}). Let us denote 
by $e_\pm^{\rm BP}(\gamma)$ the lateral Borel--Pad\'e resummations of the perturbative series, in which we use a Pad\'e approximant to extend analytically $\widehat e (\gamma)$ 
along the positive real axis (as usual in this technique, we obtain an approximate value for $e_\pm^{\rm BP}(\gamma)$ by keeping the digits which are stable as we increase 
the number of coefficients in the series). For $B=3$ one finds $\gamma(3)=0.4438...$, and 
\be
\label{lateral}
e_\pm ^{\rm BP}(\gamma(3))=0.582948901906965173...\mp 1.046566...\cdot 10^{-10} \, \ri\,. 
\ee
On the other hand, the exact ground state energy, obtained by numerical solution of the Bethe ansatz, is 
\be
e(\gamma(3))=0.58294890190696517372... 
\ee
Note that the real part of (\ref{lateral}) agrees with the exact value $e(\gamma(3))$ with a precision of eighteen digits. In addition, 
the r.h.s. of (\ref{im-ex}) for this value of $\gamma$ is $0.98... \cdot 10^{-10}$, which 
compares well with the imaginary part of (\ref{lateral}).

In order to recover the exact ground state energy, the perturbative series (\ref{eten}) should be extended to a full {\it trans-series} \cite{mmlargen,abs}, including exponentially small 
corrections in the coupling constant. The trans-series is expected to be of the form, 
\be
e(\gamma) \sim \sum_{n \ge 0} c_n \gamma^n + \sum_{\ell\ge 1} C_\ell \, \re^{- {\ell \pi^2 /\gamma}} \gamma^{b_\ell} \sum_{n \ge 0} c_n^{(\ell)} \gamma^n. 
\ee
The first term in the r.h.s. is the conventional perturbative series, while the second term 
is the infinite series of exponentially small corrections. $C_\ell$ and $b_\ell$ are parameters of the trans-series, and we note 
that the $C_\ell$ are generically complex. From the form of the ambiguity (\ref{np-amb}) it follows that $b_1=1$. We conjecture that the first coefficient has the value 
\be
\label{c1-pi}
C_1=\pm \ri 
\ee
and plays the r\^ole of a Stokes parameter. As usual, the choice of sign is correlated with the choice of lateral resummation. To justify this, we note that with the value (\ref{c1-pi}), 
the resummation of the first exponential correction to the trans-series cancels the imaginary part of the lateral Borel resummations of the perturbative series. In addition, we note that the real part of the 
Borel resummation of the perturbative series agrees with the exact result with a much higher precision than $\exp(-\pi^2/\gamma)$. Therefore, the first correction to the real part is likely to be of 
order $\exp(-2 \pi^2/\gamma)$, and this implies that the real part of $C_1$ vanishes. We note that the structure of the trans-series appearing here is very similar 
to the Painlev\'e II example studied in \cite{mmnp}.

In the {\it repulsive} case, the coefficients become alternating: $c_k \rightarrow (-1)^k c_k$. The leading singularity is now on the negative 
real axis, and the series is likely to be Borel summable. Let us denote by $e_r^{\rm BP}(\gamma)$ the Borel--Pad\'e resummation of the perturbative series. Up to the numerical accuracy with which we can 
evaluate this resummation, $e_r^{\rm BP}(\gamma)$ is in perfect agreement with the exact $e_r(\gamma)$ calculated from the Bethe ansatz. Let us give an example. 
We parametrize again $\gamma$ by $B$, as dictated by (\ref{gammaB-rep}). Then, one has, for $B=1$,  $\gamma(1)= 2.249..$, and 
\be
 e_r(\gamma(1))= 1.63146548849817237871545..., \qquad e^{\rm BP}_r(\gamma(1))=1.631465488498172....
\ee
The Borel summability of this series in the repulsive case is in agreement with general consideration on the fermion many-body perturbative expansion \cite{baker-review, baker-pirner}.

\subsection{Rings, ladders and renormalons}

Non-perturbative ambiguities in asymptotic series are the signal of non-perturbative effects that have to be taken into account. It is then important to identify the source of 
the ambiguity found in the previous section. In general, in quantum theory there are two sources for the factorial growth of the coefficients in the perturbative series (see \cite{mmlargen} for a review). 
The first one is the growth in the number of Feynman diagrams as we increase the order. 
This is believed to be encoded in the instantons of the model, i.e. in non trivial saddle points of the Euclidean path integral. 
Another source of factorial growth are renormalons. 
Renormalons are special types of diagrams whose contribution grows factorially as one increases the order. This growth is not due to an increase 
in the number of diagrams, but to the integration over momenta. Typical examples of renormalons are bubble 
diagrams in QED or QCD \cite{beneke}. 

In theories with fermions, instanton methods suggest that the growth of the perturbative series is milder than the standard factorial growth, due to 
cancellations among diagrams \cite{parisi-fermi,baker-pirner}. For example, in some models in dimension three, instanton dominance in the large order 
behavior gives a growth of order 
$(n!)^{1/5}$, instead of $n!$ \cite{baker-pirner}\footnote{In \cite{rossi}, the perturbative series for a Fermi gas in 3D with contact interaction was resummed in terms of an effective coupling, and 
a growth of $(n!)^{1/5}$ was predicted with instanton techniques. The setting is different from our since the effective coupling includes in particular the resummation of ladder diagrams. 
We would like to thank F\'elix Werner for clarifications on this issue.}. This instanton-induced growth is expected to 
become even slower as we decrease the dimension. As we have seen, the ground state energy density in the Gaudin--Yang model has a fully factorial growth, so it seems 
unlikely that this is due to an instanton effect. 

We turn then to the other possible source of factorial growth: renormalons. Are there sequences of diagrams in the Gaudin--Yang model 
which grow factorially after integration over momenta? If so, can they lead to the behavior we have 
found empirically? Before answering these questions, let us point out that the analysis of special classes of diagrams by itself does not allow to find the precise location 
of the Borel singularities. In the study of renormalons 
in quantum field theory, this is typically complemented by considerations based on the renormalization group and the OPE (see \cite{beneke}). However, the existence of these diagrams 
indicates that instanton estimates of the large order growth are likely to be inaccurate. 

In the study of renormalons in gauge theories with fermions, it has been proved useful to organize the diagrams by considering the limit of a large number of flavours. We can follow the same procedure 
here, since as we have explained there is a similar limit in many-fermion theories: the limit of a large number of spin components. This limit is dominated by the 
ring diagrams appearing in (\ref{sum-ring}). The first question to ask is then: what is the behavior of the integrals $\CI_\ell$ as $\ell$ grows large? A little bit of numerical experimentation 
indicates that indeed, they grow factorially. This can be seen as follows. Let us consider the formal power series, 
\be
\varphi(\zeta)=\sum_{\ell=2}^\infty {\zeta^\ell \over \ell!} \CI_\ell, 
\ee
which is essentially the Borel transform of the series appearing in (\ref{sum-ring}). The factorial behavior 
\be
\CI_\ell \sim C^{-\ell} \ell! 
\ee
is equivalent to the statement that the singularity of $\varphi(\zeta)$ which is closest to the origin occurs at $\zeta=C$. 
We can locate this singularity as follows. By using the explicit expression (\ref{il-int}), we resum the series defining $\varphi(\zeta)$ as
\be
\varphi(\zeta)= \int_0^\infty \rd y \, y \int_0^\infty \rd \nu \left[ f(\nu, y)^{\zeta\over 2y} - 1 -{\zeta \over 2y} \log f(\nu, y) \right], 
\ee
where
\be
 f(\nu, y)= {(y/2+1)^2 + \nu^2 \over (y/2-1)^2 + \nu^2 }. 
 \ee
The most important contribution to the above integral comes from the neighbourhood of the logarithmic singularity at $\nu=0$, $y=2$. If we introduce the variable
\be
z={y \over 2}-1,
\ee
the singularity occurs now at $z=\nu=0$. We can approximate the integral defining $\varphi(\zeta)$ by 
\be
\int_{-\Lambda_z}^{\Lambda_z} \rd z \int_0^{\Lambda_\nu} \rd \nu \left( {\nu^2 + 4 \over \nu^2 + z^2} \right)^{\zeta/4}, 
\ee
where $\Lambda_z$, $\Lambda_\nu$ define a region around the singularity. It is easy to see that this integral is finite for $0<\zeta<4$, but diverges as $\zeta\rightarrow 4$. 
The first singularity occurs at
\be
\zeta=4. 
\ee
We conclude that
\be
\CI_\ell\sim 4^{-\ell} \ell! 
\ee
and the ring diagram with $\ell$ bubbles diverges then as 
\be
\pi^{-2 \ell} \ell!. 
\ee
This corresponds to a non-perturbative ambiguity at large $\kappa$ of the form $\re^{-\pi^2/\lambda}$. When $\kappa=2$, the exponential 
dependence is precisely the one found in (\ref{np-amb}). As a side remark, let us point out that, as a consequence of this large order behavior, the series defining $e_1(\lambda)$ in (\ref{sum-ring}) is factorially divergent. This 
illustrates the fact that, in the presence of renormalons, the large $N$ expansion fails to produce convergent series order by order in $1/N$, as noted in e.g. \cite{fkw}. 

The above analysis shows that there are renormalon-type singularities in the Gaudin--Yang model: ring diagrams grow factorially, due to integration over the momenta. 
Note that the source of the growth of the integral is the logarithmic divergence of the Lindhard function at $\omega=0$, $q=2 k_F$, which is due to particle-hole excitations near 
the Fermi surface (this is in contrast to conventional renormalons in relativistic field theory, where the factorial growth comes from large or small momenta). The strength of the factorial growth is precisely what 
is needed to reproduce the observed large order behavior. 

It should be noted however that the factorial growth of the ring diagrams is specific to one dimension. 
In dimensions two and three, the Lindhard function does not have a logarithmic 
singularity, and no factorial growth of the ring diagrams is observed for a contact interaction. Is there are another set of diagrams which might lead to a factorial growth of the renormalon type? 
A natural candidate is the set of ladder diagrams shown in \figref{ladders}. The factorial growth of this class of diagrams has been noted long ago 
\cite{baker-review}, and they play an important r\^ole in the study of the Cooper instability (see e.g. \cite{as}). 

Let us then look at the pp ladder diagrams considered in section \ref{pert-subsec}. Like before, let us consider the series
\be
\label{borel-pp}
\widehat e_{\rm pp}(\zeta)= \sum_{n \ge 1} {c_n^{\rm pp} \over (n-1)!} \zeta^{n-1}, 
\ee
which is the Borel transform of (\ref{epp-coefs}). It can be written in closed form as
\be
\widehat e_{\rm pp}(\zeta)=\int_0^1 \rd s \int_0^{1-s} \rd t \, \left( {1+s -t \over 1+s + t} \right)^{-{\zeta \over t \pi^2}}. 
\ee
The most important contribution to the integral comes from $s=0$, $t=1$. This corresponds to a pair of holes with opposite momentum near the Fermi surface. 
To extract this contribution, we can approximate the integral by 
\be
\int_0^1 \rd s \int_0^{1-s} \rd t \, (1+s-t)^{-\zeta/\pi^2}, 
\ee
which has its first singularity at 
\be
\zeta=2 \pi^2. 
\ee
We deduce that the coefficients $c_n^{\rm pp}$ behave as 
\be
c_n^{\rm pp} \sim (2 \pi^2)^{-n} n!. 
\ee
This can be also confirmed by a direct numerical evaluation of the coefficients. 
In this case, the singularity associated to this sequence of diagrams occurs at $\zeta=2 \pi^2$, which is twice the value we have obtained in 
the previous section by studying the whole sequence. 
Therefore, the pp ladder diagrams underestimate the growth 
of the full sequence. This is not surprising, since we are not even taking into account the full contribution of ladder 
diagrams, but only those with two hole lines. In two \cite{kaiser2d} and three dimensions \cite{kaiser1}, it is possible to include as well 
ladders with more than one pair of hole lines. This might change the location of the Borel singularity, and as we will see in 
section \ref{sec-models}, in three dimensions it does. It would be very interesting to extend  
the calculations of \cite{kaiser1, kaiser2d} to one dimension, and check whether the inclusion of additional ladder diagrams gives the correct
singularity in the Borel plane. 

In conclusion, there are factorially divergent subdiagrams in the Gaudin--Yang model leading to the correct 
singularity (in the case of ring diagrams) or to a closely related one (in the case of pp ladder 
diagrams). In our view, this supports the argument that the non-perturbative ambiguity we have found in the 
Gaudin--Yang model is of the renormalon type.

\sectiono{More examples} 
\label{sec-models}

The connection between the non-perturbative ambiguity and the superconducting gap in weakly interacting Fermi systems is very natural, and as we have seen 
in the last section, there is clear numerical evidence that it holds in the Gaudin--Yang model. It would be interesting to investigate whether it holds in other models. In this section 
we look at two additional examples. Our first example, the one-dimensional Hubbard model at half-filling, 
is even simpler than the Gaudin--Yang model, and we can establish the conjecture analytically. The second example, the dilute Fermi gas in three dimensions, is much more difficult to analyze, but we 
can still give some partial evidence for our conjecture. 

\subsection{The Hubbard model}

The Hubbard model in one dimension describes $N$ interacting fermions of spin $1/2$ in a lattice with $L$ sites. The Hamiltonian is 
\be
\mH= -\sum_{l=1}^L \sum_{\sigma} \left( c_{l \sigma}^\dagger c_{l +1\sigma}+ c_{l+1\sigma}^\dagger c_{l \sigma} \right)
+ 2 c \sum_{l=1}^L c_{l \uparrow}^\dagger c_{l \uparrow} c_{l \downarrow}^\dagger c_{l \downarrow}, 
\ee
where $c_{l \sigma}^\dagger$, $c_{l \sigma},$ $l=1, \cdots, L$,  $\sigma= \uparrow, \, \downarrow$ are creation/annihilation operators for the 
fermions. This model can be regarded as a discretized version of the 
Gaudin--Yang model, and we will focus again on the attractive case with $c<0$.    

The ground state of the system corresponds to the so-called half-filled band, in which $N=L$, and 
with a vanishing polarization. The model can be solved with Bethe ansatz techniques \cite{lw}, and the solution leads to an exact expression for the energy of the ground state. 
In the thermodynamic limit, the ground state energy density can be written down in closed form in terms of Bessel functions:
\be
\label{e-bessel}
E=-|c| -4 \int_0^\infty {\rd \omega \over \omega} {J_0(\omega) J_1(\omega) \over 1+ \re^{|c| \omega}}. 
\ee
(We follow here the conventions in \cite{sb-book}.) This has a asymptotic series expansion in powers of $c$ which was obtained in \cite{mis-ov}. It reads, 
\be
\label{hub-ser}
E=-{4 \over \pi} -{|c| \over 2}- \sum_{k \ge 1} a_k c^{2k}, 
\ee
where
\be
a_k ={ (2k-1) (2^{2k+1}-1) ( (2k-3)!!)^3 \over 2^{5 k-3}(k-1)!} {\zeta (2k+1) \over \pi^{2k+1}}. 
\ee
The very first terms in this series can be tested against conventional perturbation theory \cite{metzner}, as we have done for the Gaudin--Yang model in previous sections. 
Using that $\zeta(x) \rightarrow 1$ as $x\rightarrow \infty$, one can determine analytically the large order behavior of the $a_k$, 
\be
\label{lob-ak}
a_k \sim (2 \pi)^{-2k+1} \Gamma(2k-1). 
\ee
The series diverges doubly-factorially and it is non-alternating, therefore non-Borel summable. The large order behavior (\ref{lob-ak}) of the $a_k$ leads to a non-perturbative ambiguity of order
\be
\label{hub-npa}
 \re^{-2 \pi/|c|}. 
\ee
On the other hand, the Hubbard model is known to have an energy gap which, in the half-filled case, is of the form (see e.g. \cite{marsiglio,giamarchi})
\be
\Delta \sim \re^{-\pi/|c|}. 
\ee
As before, the ambiguity (\ref{hub-npa}) has the exponential dependence of the square of the gap. It would be very interesting to understand the origin of the Borel singularity by looking at the 
diagrammatic structure of the theory, as we have done for the Gaudin--Yang model. 

\subsection{The dilute Fermi gas in three dimensions}

The two models discussed so far are the Gaudin--Yang model and the one-dimensional Hubbard model. They are both integrable and one can obtain explicit expressions 
for the coefficients of the perturbative series to large order (in the case of the Hubbard model with a half-filled band, to all orders). It would be interesting to explore other models 
where our conjecture can be tested, in particular models in higher dimensions, although an important limitation in this study is the lack of concrete results on the coefficients of the perturbative 
series. 

As a modest first step in this direction, we will make a preliminary analysis of a dilute Fermi gas in three dimensions with a contact interaction. The Lagrangian of the system is 
written as
\be
\CL= \psi^\dagger\left( \ri \partial_t + {1\over 2M } \nabla^2 \right) \psi -{C_0 \over 2} (\psi^\dagger \psi)^2,
\ee
where $M$ is the mass of the fermions. 
One can include additional interactions in order to match the scattering data. In this matching, the constant $C_0$ is related to the scattering length $a$ by $C= 4 \pi a/M$. The natural 
dimensionless coupling is
\be
g=k_F a, 
\ee
where $k_F$ is the Fermi momentum. We follow the convention 
in \cite{kaiser1}, namely, an attractive interaction corresponds to $g>0$. One can expand the ground state energy per particle in a power series in $g$. Note that, in contrast to the Gaudin--Yang model, the weak coupling limit of 
small $g$ is here a limit of low density. Let us define 
\be
e^{\rm 3D}(g)= {2 M \over k_F^2} {E \over N}, 
\ee
where $N$ is the number of particles. Then, after including the diagrams shown in \figref{pert-series}, one finds the well-known 
result (see e.g. \cite{bishop,hf-effective}), 
\be
\label{e3}
e^{\rm 3D}(g)={3\over 5} -{2 g \over 3 \pi} +{4\over 35 \pi^2} \left(11- 2 \log(2)\right) g^2+ \CO(g^3), 
\ee
where we have set for simplicity $\kappa=2$. This series is the three-dimensional counterpart to (\ref{eten}), but it is only known up to order $g^4$ \cite{fourth-order}. 
An additional subtlety as compared to (\ref{eten}) is that $e^{\rm 3D} (g)$ also includes logarithmic corrections (i.e. terms involving $\log (g)$), so one should enlarge the formal framework 
we have used in the Gaudin--Yang model to take into account this additional expansion parameter. However, we can have a first diagnose on the 
large order behavior of (\ref{e3}) by looking at renormalon effects similar to those unveiled in the Gaudin--Yang model. In three dimensions, ring diagrams grow only exponentially, 
so we should look at ladder diagrams. As in section \ref{pert-subsec}, we first consider pp ladder diagrams, which contain one hole-hole line and $n$ particle-particle lines. 
They can be resummed in a compact form (see e.g. \cite{steele,schafer, kaiser1,he-huang}), and the result reads
\be
e^{\rm 3D}_{\rm pp}(g) = {3\over 5}- {48 g \over \pi} \int_0^1 \rd s \, s^2 \int_0^{\sqrt{1-s^2}} \rd t \, t \, { I(s, t) \over 1+ {g \over \pi} F_{\rm pp}(s, t)}. 
\ee
Here, the function $F_{\rm pp}(s, t)$ is the three-dimensional analogue of (\ref{pp1}), and it reads 
\be
 F_{\rm pp}(s, t)=1+s + {1- t^2-s^2 \over 2s} \log \left( {(s+1)^2-t^2 \over 1-t^2-s^2} \right) 
+ t \log \left(\frac{t+s+1}{\left|1-t+s\right| }\right). 
   \ee
 The function $I(s,t)$ was introduced in \cite{kaiser2} and it is a kinematic function to perform the integration over hole momenta. Its explicit expression is 
\be
I(s, t)= \begin{cases} t, & \text{for $0<t<1-s$}, \\
{1\over 2s}(1-s^2 -t^2), & \text{for $1-s<t< {\sqrt{1-s^2}}$}.
\end{cases}
\ee
We can now re-expand $e^{\rm 3D}_{\rm pp}(g)$ in powers of $g$, to obtain
\be
e^{\rm 3D}_{\rm pp}(g)= \sum_{n \ge 0} c_n^{\rm 3D, \, pp}g^n, 
\ee
 where
 \be
 c^{\rm 3D, \, pp}_n = - {48 \over \pi} \int_0^1 \rd s \, s^2 \int_0^{\sqrt{1-s^2}} \rd t \, t \,  I(s, t)  \left( -{F_{\rm pp}(s, t) \over \pi} \right)^{n-1}, \qquad n \ge 1. 
 \ee

It is also possible to include and resum ladders with more than one pair of hole lines \cite{kaiser1, kaiser2d} (see also \cite{he}). One finds, 
\be
e^{\rm 3D}_{\rm lad}(g) = {3\over 5}- {48 \over \pi} \int_0^1 \rd s \, s^2 \int_0^{\sqrt{1-s^2}} \rd t \, t \, \arctan \left[ { I(s, t) g \over 1+ {g \over \pi} R(s, t)} \right], 
\ee
where the function $R(s, t)$ is given by 
\be
 R(s, t)=F_{\rm pp}(s,t)+  F_{\rm pp}(-s,t). 
 \ee
After re-expanding, we find 
\be
e^{\rm 3D}_{\rm lad}(g)= \sum_{n \ge 0} c_n^{\rm 3D, \, lad}g^n. 
\ee

\begin{figure}[tb]
\begin{center}
\begin{tabular}{cc}
\resizebox{60mm}{!}{\includegraphics{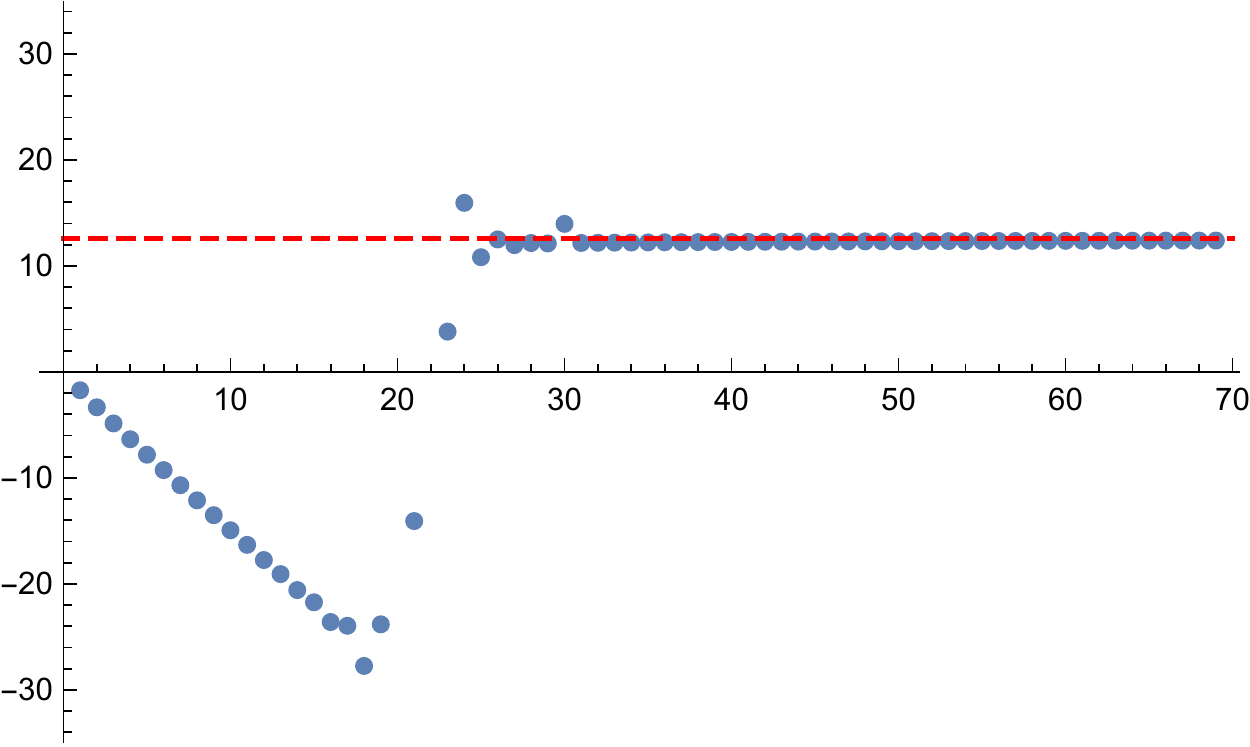}}
\hspace{3mm}
&
\resizebox{60mm}{!}{\includegraphics{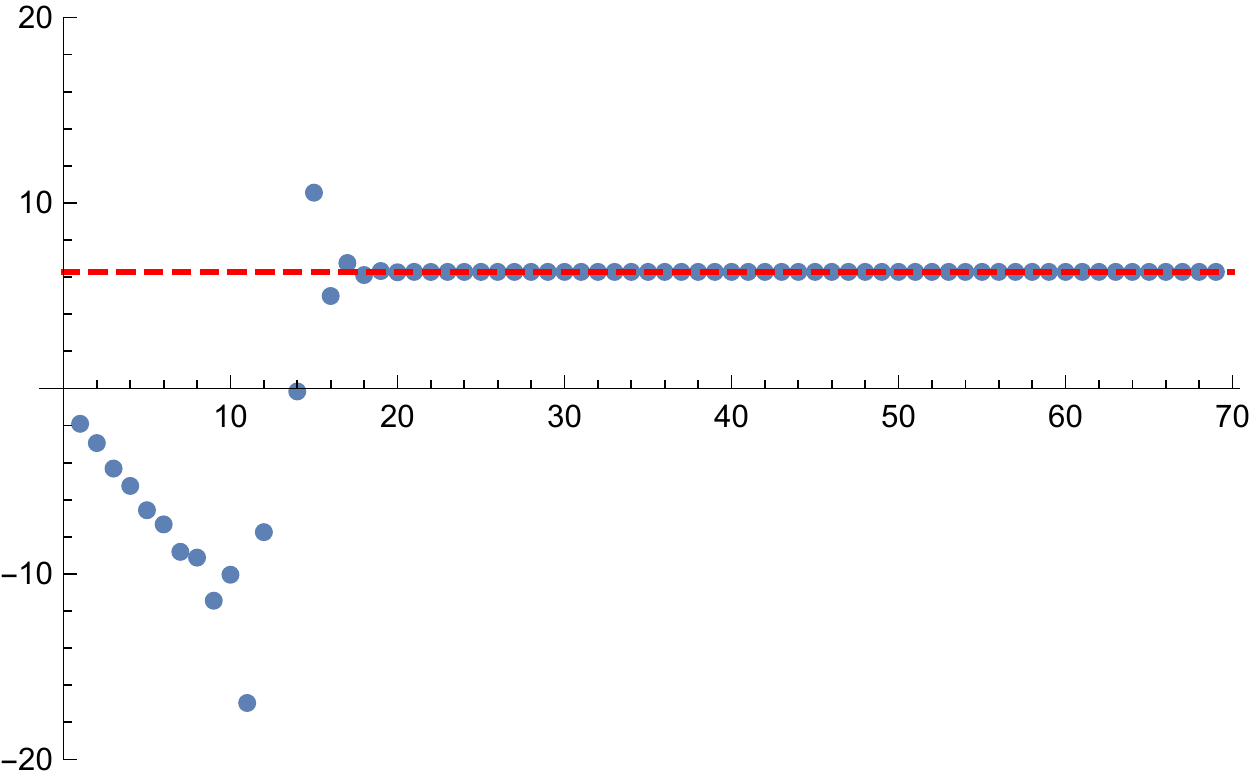}}
\end{tabular}
\end{center}
  \caption{The sequence $n c_n^{\rm 3D, \, pp}/c_{n+1}^{\rm 3D, \, pp}$ (left) and  $n c_n^{\rm 3D, \, lad}/c_{n+1}^{\rm 3D, \, lad}$ (right). 
The dashed lines are, respectively, $4 \pi$ and $2 \pi$. As we can see, the asymptotic behavior only manifests itself when $n \gtrsim 25$ and $n \gtrsim 15$, 
respectively. 
}
\label{3dladders}
\end{figure}

We can now ask what is the large order behavior of the coefficients $c_n^{\rm 3D, \, pp}$, $c_n^{\rm 3D, \, lad}$. This question can be addressed either with the 
Borel resummation techniques we used for the ring and ladder 
diagrams in the Gaudin--Yang model, or by inspecting numerically the behavior of the coefficients at large $n$. One finds, 
\be
\label{ppph-growth}
c_n^{\rm 3D, \, pp} \sim (4 \pi)^{-n} n!, \qquad c_n^{\rm 3D, \, lad} \sim (2 \pi)^{-n} n!\, . 
\ee
This can be clearly seen in the plots in \figref{3dladders}, although the actual large order behavior sets in relatively late. 
In the pp case, for example, the series seems to grow exponentially (and to be alternating) until $n \sim 25$. 

The first conclusion to extract from (\ref{ppph-growth}) is that ladder diagrams indeed grow factorially and lead to non-Borel summable series. 
The second conclusion is that, as we enlarge the type of diagrams included in the resummation, the location of the first Borel singularity 
changes as well: the pp ladders lead to a singularity at $4 \pi$, while the improved resummation in 
\cite{kaiser1} moves it to $2 \pi$. 
The associated non-perturbative ambiguities are 
\be
\text{pp}: \quad \re^{-{4 \pi \over k_F a}}, \qquad \text{lad}: \quad \re^{-{2 \pi \over k_F a}}. 
\ee
We can now compare this to the result for the superconducting gap, which is of the form (see e.g. \cite{pitaevski-rev})
\be
\Delta_{\rm BCS} \sim \re^{-{\pi \over 2 k_F a}}. 
\ee
According to our conjecture, the perturbative series should lead to a Borel singularity at $\pi$, which is {\it half} the singularity obtained with the improved ladder resummation of \cite{kaiser1}. 
This apparent mismatch shouldn't be discouraging, since our conjecture concerns the {\it full} series, while the results above concern only a subclass of diagrams. 
Indeed, the 
location of the singularity changes 
by a factor of two as we pass from pp ladders to ladders including at least two holes lines. It is conceivable that including more diagrams would confirm our conjecture. In addition, one should remember that the 
perturbative series for the three-dimensional gas includes logarithms, so probably more care should be exercised. We find it 
encouraging that well-known resummations of the ground state energy of the dilute Fermi gas lead to Borel singularities which are closely related to the superconducting energy gap. In any case, a more detailed investigation is needed 
in order to determine the large order behavior of the ground state energy for this system.  

\sectiono{Conclusions}

\label{sec-conclude}

The main goal of this paper has been to start the study of perturbative series in many-body physics with the tools of the theory of resurgence. 
History teaches us that a useful starting point is a non-trivial, yet quasi-solvable model where one can obtain extensive data on its perturbative structure. 
This is what Bender and Wu did in quantum mechanics: they first obtained data for the perturbative series of the ground state energy of the quartic oscillator \cite{bw}, and 
then they discovered its precise connection to tunneling \cite{bw2}. 

In this paper we have tried to adapt this program to many-body fermionic systems. Our toy model has been the 
Gaudin--Yang model, which can be solved exactly with the Bethe ansatz. However, extracting the perturbative series 
for the ground state energy from the exact solution is already non-trivial. Fortunately, we were able to rely on the powerful techniques of \cite{volin, volin-thesis}, and we found a systematic method to obtain 
the perturbative information. The series turns out to be factorially divergent and non-Borel summable, and we have shown that its large order behavior is controlled by the 
superconducting energy gap of the model. 

In retrospect, the validity of our conjecture is very natural in view of what happens in the Gross--Neveu model \cite{gross-neveu}, which can be regarded as a relativistic cousin of the attractive 
Gaudin--Yang model. At large $N$, which corresponds to the BCS mean-field theory, the mass gap of the Gross--Neveu model can be calculated exactly and is of the form \cite{fnw2}
\be
m \sim \re^{-{1\over 2 \beta_0 g^2}}, 
\ee
where $\beta_0$ is the first coefficient of the beta function. On the other hand, this non-perturbative effect controls the large order behavior of the {\it conventional}
perturbative series in the model, and it is a renormalon effect. Our conjecture is a transposition of these well-known facts to the non-relativistic case. 

The energy gap is a robust non-perturbative effect in weakly coupled Fermi systems with an attractive interaction, and it sets the 
non-perturbative scale of the problem. Therefore it is natural to 
conjecture that this connection is general, and that the resurgent structure of superconducting systems is governed by the superconducting gap. We have also 
stated a quantitatively precise connection, in (\ref{en-gap}), which holds in the Gaudin--Yang model and in the closely related example of the one-dimensional Hubbard model. 
Our conjecture should be regarded as an exploration tool in a research program, so it might need qualification or modification in more general cases. Our preliminary 
analysis of the dilute Fermi gas in three dimensions indicates that the large order behavior of special types of diagrams goes into the right direction, but more work is clearly necessary. Recent progress 
on high-order calculations (as in e.g. \cite{montecarlo}) might shed light on these issues.  

Our results seem to provide an interesting picture of the Cooper instability. It is often stated that, due to this instability, 
the perturbative expansion around the conventional vacuum is incorrect, and one should expand around 
the superconducting vacuum with a non-zero gap. Our findings make this statement more precise: the conventional 
perturbative series is a non-Borel summable series, which lacks non-perturbative information. In quantum theory, non-Borel summable 
series are often a signal that one is not expanding around the right ground state, or that one is not taken into account additional vacua. 
At the same time, the theory of resurgence teaches us that the shortcomings of conventional perturbation theory can be cured in many cases
by extending the perturbative series to a trans-series. Therefore, 
once the full trans-series for the superconductor is obtained, it should be in principle possible to extract its physical 
ground state energy. 

We have also argued that the non-perturbative ambiguity associated to the lack of Borel 
summability does not have a semiclassical interpretation, and is due to 
a renormalon-like effect. To this end, we have identified sequences of diagrams which 
display the typical factorial growth of renormalon physics, coming from integration over momenta. In the case of ladder diagrams, 
which are known to encode information about the Cooper instability, the kinematic region that leads to the factorial growth corresponds indeed 
to Cooper pairs near the Fermi surface. 

The connection between the Cooper instability and 
renormalon singularities leads to a very intriguing interplay, and should be understood further\footnote{Some parallelisms between 
superconductivity and the large order behavior of perturbative 
series in QCD were pointed out in \cite{peris}. Renormalon-like diagrams were found in the $\epsilon$ expansion of the unitary Fermi gas in \cite{nishida-son}.}. 
It would be interesting to see whether standard tools in renormalon 
physics, like the use of RG equations and the OPE, can be adapted to many-body physics and shed light on the location of the Borel singularities. 
This interplay might also lead to a more tractable arena to understand the phenomenon of 
renormalons in quantum field theory. It has been argued in recent years that renormalons have semiclassical realizations once the theory is 
suitable compactified \cite{argyres-unsal, a-unsal-long, dunne-unsal-cpn, cherman-dorigoni-dunne-unsal, cdu, misumi-1,du-on, misumi-2}. It is not clear whether this scenario 
holds in purely fermionic theories. For example, the Gross--Neveu model, which is historically the first 
model displaying renormalons \cite{gross-neveu}, has not been given so far a semiclassical picture. It would be then interesting 
to see whether the Borel singularity we have found in the Gaudin--Yang model can be interpreted in semiclassical terms along the lines proposed in those works. 

In addition to these more general issues, it would be fruitful to apply the techniques used in this paper to other models. One 
could for example study the Gaudin--Yang model with arbitrary polarization, for which some terms in the series are known \cite{iw-pol,guan-review}. More generally, one can apply  
the method developed in this paper to other integrable theories in two dimensions \cite{mr-ift}. The techniques we have developed can be also applied to the Lieb--Liniger model \cite{mr-ll}, and one obtains 
in this way explicit results for the perturbative series of its ground state energy density. 
The series obtained in that case is also factorially divergent and non-Borel summable. Given the importance of the Lieb--Liniger model, both theoretically and experimentally, 
it would be very interesting to have a precise physical understanding of the corresponding non-perturbative ambiguity. It is likely that this is due to an instanton 
in the non-relativistic version of the $O(2)$, two-dimensional $\lambda \phi^4$ theory describing the Lieb--Liniger model.

Another important question is to obtain the precise form of the trans-series for the Gaudin--Yang model. 
This trans-series is encoded in the exact Bethe ansatz solution, 
but it is not obvious how to extract it. Perhaps one should first address this question in the half-filled, one-dimensional Hubbard model, where the 
energy can be written down in closed form, as we have seen in (\ref{e-bessel}). A 
closely related question would be to find the all-orders expansion of the spin gap of the Gaudin--Yang model from the Bethe ansatz. 
The leading order term quoted in (\ref{spin-gap}) was determined in \cite{ko} up 
to an overall constant, which was fixed in \cite{frz} by using 
results from the Hubbard model\footnote{We would like to thank W. Zwerger for detailed explanations of this point.}. It would be 
very interesting to see whether the techniques used in this paper to study the ground state energy can be also used to calculate the subleading corrections to the energy gap.

More generally, it would be important to explore with the tools of resurgence the vast landscape of models of interacting fermions, and to study the connection conjectured in this paper 
between Borel singularities and the non-perturbative energy gap. This might lead to a deeper understanding of non-perturbative effects in these theories.

\section*{Acknowledgements}
We would like to thank Lianyi He, Norbert Kaiser, Sylvain Prolhac, Ricardo Schiappa, and Wilhelm Zwerger for useful discussions and correspondence. We are specially indebted to Thierry Giamarchi for 
his patient explanations of one-dimensional models, and to F\'elix Werner for a detailed reading of the manuscript. This work is is supported in part by the Fonds National Suisse, 
subsidies 200021-175539, and by the NCCR 51NF40-182902 ``The Mathematics of Physics'' (SwissMAP).

\appendix

\sectiono{Perturbative evaluation of the NNNLO contribution}
\label{3-diagrams}

The contribution to the ground state energy of the Gaudin--Yang model at order $\gamma^3$ was obtained in \cite{magyar2} by using standard perturbative techniques. Here we rederive 
this result and we consider an arbitrary number of spin components $\kappa$. We can adapt the results of \cite{bishop}, who studied the perturbative expansion of the dilute Fermi gas in three 
dimensions. At order $\gamma^3$ there are three types of Goldstone 
diagrams that contribute: a purely pp ladder diagram (together with its exchange), a ladder diagram with 2 particle lines and 4 hole lines, 
and finally diagrams with 3 particle lines and 3 hole lines (which include 
the three-bubble ring diagram). The total contribution is of the form 
\be
-{\gamma^3 \over 2 \pi^4} \left( \kappa(\kappa-1) (I_1+ I_2) + \kappa (\kappa-1) (\kappa-3) I_3\right),
\ee
where $I_j$, $j=1,2,3$ correspond to the types of diagrams that we have just listed. $I_1$ comes from the pp ladder, which according to (\ref{e-pp-ladder}) is given by 
\be
I_1=  \int_0^1 \rd s \int_0^{1-s} \rd t \left(\frac{1}{t}\log\left(\frac{1+s+t}{1+s-t}\right)\right)^2= \pi^2 \log(2)-\frac{9}{2}\zeta(3). 
\ee
$I_2$ is given by 
\be
I_2=\frac{1}{2}\int_{\IR^2} \rd m\,\rd m'\,(1-n(m))(1-n(m'))\left(\int_{\IR^2} \frac{n(p)n(p') \rd p\, \rd p' }{p^2+p'^2-m^2-m'^2}\delta(p+p'-m-m')\right)^2, 
\ee
where $n(k)=\theta(1-|k|)$. After some manipulations, we obtain 
\be
I_2= 4 \int_0^{1} \rd s \int_{1+s}^{\infty} \rd t \left(\frac{1}{2t}\log\left(\frac{t-s+1}{t-1+s}\right)\right)^2=-\pi^2\log(2)+6\zeta(3). 
\ee
Finally, we have 
\be
I_3= \int_{\IR^2} \rd p\,\rd m (1-n(m))n(p) \left(\int_{\IR^2} \rd q\, \rd q'\frac{(1-n(q'))n(q)}{p^2+q^2-m^2-q'^2}\right)^2. 
\ee
It is convenient so split this integral in four parts, as follows, 
\be
\ba
8 I_3&= \int_{-\infty}^{-1} \rd t \int_{t-1}^{t+1} \rd s\left(\frac{1}{t}\log\left(\frac{t+s+1}{t+s-1}\right)\right)^2+ \int_{-1}^{0} \rd t \int_{t-1}^{-t-1} \rd s\left(\frac{1}{t}\log\left(\frac{-t+s-1}{t+s-1}\right)\right)^2\\
&+\int_{-1}^{0} \rd t \int_{-t+1}^{t+1}  \rd s\left(\frac{1}{t}\log\left(\frac{-t+s+1}{t+s+1}\right)\right)^2+\int_{1}^{\infty} \rd t \int_{t-1}^{t+1} \rd s\left(\frac{1}{t}\log\left(\frac{t+s+1}{t+s-1}\right)\right)^2. 
\ea
\ee
This can be evaluated as 
\be
I_3= {\zeta(3) \over 2}. 
\ee
Since
\be
I_1+ I_2-I_3=\zeta(3), 
\ee
for $\kappa=2$ we obtain the result (\ref{c3}). This confirms the Bethe ansatz calculation of the term of order $\gamma^3$ in (\ref{eten}). In addition, since we also have the 
explicit spin factors, we can also verify that the contribution to $e(\lambda; \kappa)$ 
of order $\lambda^3$ is
\be
-{4 \over \pi^4} \zeta(3) \kappa^{-1}+ {4 \over \pi^4} \zeta(3) \kappa^{-2}, 
\ee
in agreement with the results for $e_{1,2} (\lambda)$ in (\ref{kappa-results}).

\sectiono{Matching the bulk to the edge}
\label{match}

Let us illustrate the method developed in section \ref{resolvent-section} by working out in detail how the matching of bulk and edge is done. By expanding (\ref{solgy}) around the edge regime, we find
 \be
 \ba
R(z)&=-\log(z)+\frac{c_{1,0,0}-\log(B\pi)c_{1,0,1}+c_{1,0,1}\log\left(\frac{\pi z}{4}\right)}{z}\\
&+\frac{1}{z^2}\biggl(c_{2,0,0}-\log(B\pi)c_{2,0,1}+\log(B\pi)^2c_{2,0,2}+\left(c_{2,0,1}-2\log(B\pi)c_{2,0,2}\right)\log\left(\frac{\pi z}{4}\right)\\
& \qquad  \qquad +c_{2,0,2}\log\left(\frac{\pi z}{4}\right)^2\biggr)-\frac{c_{1,0,1}\log(z)}{4B}\\
&+\frac{1}{zB}\left(c_{1,1,0}-\frac{1}{4}c_{2,0,1}-\log(B\pi)c_{1,1,1}+\frac{\log(B\pi)}{2}c_{2,0,1}+\frac{\log(B\pi)}{2}c_{2,0,2}+\log(B\pi)^2c_{1,1,2}\right)\\
& +\frac{\gyl}{zB}\left(c_{1,1,1}-\frac{1}{2}c_{2,0,1}-\frac{1}{2}c_{2,0,2}-2\log(B\pi)c_{1,1,2}\right)+\frac{c_{1,1,2}\gyl^2}{z}+\mathcal{O}\left(\frac{1}{B^2}\right).\\
\ea
\ee
On the other hand, the expansion of (\ref{sol1}) at $s=0$ reads, 
\be
\ba
\hat{R}(s)&=\frac{1}{s}-\frac{-1+\gamma_E+\log\left(\frac{4 s}{\pi}\right)}{\pi}\\
&+\frac{s}{2\pi^2}\left((-1+\gamma_E)^2+\frac{\pi^2}{2}+\log\left(\frac{4 s}{\pi}\right)\left(-2+2\gamma_E+\log\left(\frac{4 s}{\pi}\right)\right)\right)\\
&+\frac{1}{{\sqrt{\pi}} B}\left(\frac{Q_{0,0}}{s}-\frac{Q_{0,0}\left(-1+\gamma_E+\log\left(\frac{4 s}{\pi}\right)\right)}{\pi}\right)+\cdots
\ea
\ee
where $\gamma_E$ is the Euler--Mascheroni constant. Its Laplace transform gives
\be
\label{fg-edge}
\ba
R(z)&=-\log(z)+\frac{1+\log\left(\frac{\pi z}{4}\right)}{\pi z}+\frac{1}{2\pi^2 z^2}\left(-1+\frac{2\pi^2}{3}+\log^2\left(\frac{\pi z}{4}\right)\right)\\
&+\frac{Q_{0,0}}{{\sqrt{\pi}} B}\left(-\log(z)+\frac{1+\log\left(\frac{\pi z}{4}\right)}{\pi z}\right)+\cdots
\ea
\ee
By comparing the coefficients we obtain,
\be
\label{res-coefs}
\ba
c_{1,0,1}&=\frac{1}{\pi}, \qquad 
c_{1,0,0}=\frac{1+\log(B\pi)}{\pi}, \qquad 
c_{2,0,2}=\frac{1}{2\pi^2}, \\
c_{2,0,1}&=\frac{\log(B\pi)}{\pi^2}, \qquad 
c_{2,0,0}=-\frac{1}{2\pi^2}+\frac{1}{3}+\frac{\log(B\pi)}{2\pi^2}, \qquad Q_{0,0}=\frac{1}{4 \sqrt{\pi}}, \\
c_{1,1,2}&=0, \qquad c_{1,1,1}=\frac{1}{2\pi^2}+\frac{\log(B\pi)}{2\pi^2}, \qquad 
c_{1,1,0}=\frac{1}{4\pi^2}+\frac{\log(B\pi)}{2\pi^2}. 
\ea
\ee

\bibliographystyle{JHEP}

\linespread{0.6}
\bibliography{biblio-1d}

\providecommand{\href}[2]{#2}\begingroup\raggedright\begin{thebibliography}{10}

\bibitem{bcs}
J.~Bardeen, L.~N. Cooper and J.~R. Schrieffer, \emph{{Theory of
  superconductivity}},
  \href{http://dx.doi.org/10.1103/PhysRev.108.1175}{\emph{Phys. Rev.} {\bf 108}
  (1957) 1175--1204}.

\bibitem{bw}
C.~M. Bender and T.~T. Wu, \emph{{Anharmonic oscillator}},
  \href{http://dx.doi.org/10.1103/PhysRev.184.1231}{\emph{Phys. Rev.} {\bf 184}
  (1969) 1231--1260}.

\bibitem{bw2}
C.~M. Bender and T.~T. Wu, \emph{{Anharmonic oscillator. 2: A Study of
  perturbation theory in large order}},
  \href{http://dx.doi.org/10.1103/PhysRevD.7.1620}{\emph{Phys. Rev.} {\bf D7}
  (1973) 1620--1636}.

\bibitem{mmlargen}
M.~Mari{\~n}o, \emph{{Lectures on non-perturbative effects in large $N$ gauge
  theories, matrix models and strings}},
  \href{http://dx.doi.org/10.1002/prop.201400005}{\emph{Fortsch. Phys.} {\bf
  62} (2014) 455--540}, [\href{http://arxiv.org/abs/1206.6272}{{\tt
  1206.6272}}].

\bibitem{abs}
I.~Aniceto, G.~Basar and R.~Schiappa, \emph{{A Primer on Resurgent Transseries
  and Their Asymptotics}},  \href{http://arxiv.org/abs/1802.10441}{{\tt
  1802.10441}}.

\bibitem{baker-review}
G.~A. Baker, \emph{Singularity structure of the perturbation series for the
  ground-state energy of a many-fermion system},
  \href{http://dx.doi.org/10.1103/RevModPhys.43.479}{\emph{Rev. Mod. Phys.}
  {\bf 43} (1971) 479--531}.

\bibitem{jw}
A.~Jackson and T.~Wettig, \emph{Necessary conditions for microscopic many-body
  theories},
  \href{http://dx.doi.org/https://doi.org/10.1016/0370-1573(94)90042-6}{\emph{Phys.
  Rept.} {\bf 237} (1994) 325--355}.

\bibitem{parisi-fermi}
G.~Parisi, \emph{{Asymptotic Estimates in Perturbation Theory with Fermions}},
  \href{http://dx.doi.org/10.1016/0370-2693(77)90020-X}{\emph{Phys. Lett.} {\bf
  66B} (1977) 382--384}.

\bibitem{baker-pirner}
G.~A. Baker, Jr. and H.~J. Pirner, \emph{{Asymptotic estimate of large orders
  in perturbation theory for the many-fermion ground state energy}},
  \href{http://dx.doi.org/10.1016/0003-4916(83)90334-2}{\emph{Annals Phys.}
  {\bf 148} (1983) 168}.

\bibitem{gaudin}
M.~Gaudin, \emph{Un systeme \`a une dimension de fermions en interaction},
  \href{http://dx.doi.org/https://doi.org/10.1016/0375-9601(67)90193-4}{\emph{Phys.
  Lett.} {\bf A24} (1967) 55 -- 56}.

\bibitem{yang}
C.-N. Yang, \emph{{Some exact results for the many body problems in one
  dimension with repulsive delta function interaction}},
  \href{http://dx.doi.org/10.1103/PhysRevLett.19.1312}{\emph{Phys. Rev. Lett.}
  {\bf 19} (1967) 1312--1314}.

\bibitem{ko}
V.~Y. Krivnov and A.~Ovchinnikov, \emph{One-dimensional {F}ermi gas with
  attraction between the electrons}, {\emph{J. Exp. Theor. Phys.} {\bf 40}
  (1975) 781}.

\bibitem{giamarchi}
T.~Giamarchi, \emph{Quantum physics in one dimension}.
\newblock Oxford University Press, 2004.

\bibitem{ll}
E.~H. Lieb and W.~Liniger, \emph{{Exact analysis of an interacting Bose gas. 1.
  The General solution and the ground state}},
  \href{http://dx.doi.org/10.1103/PhysRev.130.1605}{\emph{Phys. Rev.} {\bf 130}
  (1963) 1605--1616}.

\bibitem{iw-gy}
T.~Iida and M.~Wadati, \emph{Exact analysis of $\delta$-function attractive
  fermions and repulsive bosons in one-dimension}, {\emph{J. Phys. Soc. Jpn.}
  {\bf 74} (2005) 1724--1736}.

\bibitem{tw1}
C.~A. Tracy and H.~Widom, \emph{On the ground state energy of the
  delta-function {F}ermi gas},
  \href{http://dx.doi.org/10.1063/1.4964252}{\emph{J. Math. Phys.} {\bf 57}
  (2016) 103301}.

\bibitem{tw2}
C.~A. Tracy and H.~Widom, \emph{On the ground state energy of the
  delta-function {F}ermi gas {II}: Further asymptotics},  in \emph{Geometric
  Methods in Physics} (P.~Kielanowski, A.~Odzijewicz and E.~Previato, eds.),
  pp.~201--212, Springer, 2018.

\bibitem{guan-review}
X.-W. Guan, M.~T. Batchelor and C.~Lee, \emph{Fermi gases in one dimension:
  From {B}ethe ansatz to experiments},
  \href{http://dx.doi.org/10.1103/RevModPhys.85.1633}{\emph{Rev. Mod. Phys.}
  {\bf 85} (2013) 1633--1691}.

\bibitem{prolhac}
S.~Prolhac, \emph{{Ground state energy of the $delta$-Bose and Fermi gas at
  weak coupling from double extrapolation}},
  \href{http://dx.doi.org/10.1088/1751-8121/aa5e00}{\emph{J. Phys.} {\bf A50}
  (2017) 144001}.

\bibitem{hutson}
V.~Hutson, \emph{The circular plate condenser at small separations},
  \href{http://dx.doi.org/10.1017/S0305004100002152}{\emph{Math. Proc.
  Cambridge Philos. Soc.} {\bf 59} (1963) 211--224}.

\bibitem{popov}
V.~N. Popov, \emph{Theory of one-dimensional {B}ose gas with point
  interaction}, {\emph{Theor. Math. Phys.} {\bf 30} (1977) 222--226}.

\bibitem{volin}
D.~Volin, \emph{{From the mass gap in O(N) to the non-Borel-summability in O(3)
  and O(4) sigma-models}},
  \href{http://dx.doi.org/10.1103/PhysRevD.81.105008}{\emph{Phys. Rev.} {\bf
  D81} (2010) 105008}, [\href{http://arxiv.org/abs/0904.2744}{{\tt
  0904.2744}}].

\bibitem{volin-thesis}
D.~Volin, \emph{{Quantum integrability and functional equations: Applications
  to the spectral problem of AdS/CFT and two-dimensional sigma models}},
  \href{http://dx.doi.org/10.1088/1751-8113/44/12/124003}{\emph{J. Phys.} {\bf
  A44} (2011) 124003}, [\href{http://arxiv.org/abs/1003.4725}{{\tt
  1003.4725}}].

\bibitem{mmbook}
M.~Mari{\~n}o, \emph{Instantons and large $N$. An introduction to
  non-perturbative methods in quantum field theory}.
\newblock Cambridge University Press, 2015.

\bibitem{beneke}
M.~Beneke, \emph{{Renormalons}},
  \href{http://dx.doi.org/10.1016/S0370-1573(98)00130-6}{\emph{Phys. Rept.}
  {\bf 317} (1999) 1--142}, [\href{http://arxiv.org/abs/hep-ph/9807443}{{\tt
  hep-ph/9807443}}].

\bibitem{lw}
E.~H. Lieb and F.~Y. Wu, \emph{{Absence of Mott transition in an exact solution
  of the short-range, one-band model in one dimension}},
  \href{http://dx.doi.org/10.1103/PhysRevLett.21.192.2,
  10.1103/PhysRevLett.20.1445}{\emph{Phys. Rev. Lett.} {\bf 20} (1968)
  1445--1448}.

\bibitem{magyar}
R.~J. Magyar and K.~Burke, \emph{Density-functional theory in one dimension for
  contact-interacting fermions},
  \href{http://dx.doi.org/10.1103/PhysRevA.70.032508}{\emph{Phys. Rev. A} {\bf
  70} (2004) 032508}.

\bibitem{magyar2}
R.~J. Magyar, \emph{Ground and excited-state fermions in a one-dimensional
  double-well: Exact and density-functional solutions},
  \href{http://dx.doi.org/10.1103/PhysRevB.79.195127}{\emph{Phys. Rev. B} {\bf
  79} (2009) 195127}.

\bibitem{steele}
J.~V. Steele, \emph{{Effective field theory power counting at finite density}},
   \href{http://arxiv.org/abs/nucl-th/0010066}{{\tt nucl-th/0010066}}.

\bibitem{hf}
R.~J. Furnstahl and H.~W. Hammer, \emph{{Effective field theory for Fermi
  systems in a large $N$ expansion}},
  \href{http://dx.doi.org/10.1006/aphy.2002.6313}{\emph{Annals Phys.} {\bf 302}
  (2002) 206--228}, [\href{http://arxiv.org/abs/nucl-th/0208058}{{\tt
  nucl-th/0208058}}].

\bibitem{schafer}
T.~Sch\"afer, C.-W. Kao and S.~R. Cotanch, \emph{{Many body methods and
  effective field theory}},
  \href{http://dx.doi.org/10.1016/j.nuclphysa.2005.08.006}{\emph{Nucl. Phys.}
  {\bf A762} (2005) 82--101}, [\href{http://arxiv.org/abs/nucl-th/0504088}{{\tt
  nucl-th/0504088}}].

\bibitem{kaiser1}
N.~Kaiser, \emph{{Resummation of fermionic in-medium ladder diagrams to all
  orders}},
  \href{http://dx.doi.org/10.1016/j.nuclphysa.2011.05.005}{\emph{Nucl. Phys.}
  {\bf A860} (2011) 41--55}, [\href{http://arxiv.org/abs/1102.2154}{{\tt
  1102.2154}}].

\bibitem{kaiser2}
N.~Kaiser, \emph{{Particle-hole ring diagrams for fermions in two dimensions}},
  \href{http://dx.doi.org/10.1016/j.aop.2014.07.045}{\emph{Annals Phys.} {\bf
  350} (2014) 549}, [\href{http://arxiv.org/abs/1408.6707}{{\tt 1408.6707}}].

\bibitem{kaiser2d}
N.~Kaiser, \emph{{Single-particle potential from resummed ladder diagrams}},
  \href{http://dx.doi.org/10.1140/epja/i2013-13140-6}{\emph{Eur. Phys. J.} {\bf
  A49} (2013) 140}, [\href{http://arxiv.org/abs/1305.6234}{{\tt 1305.6234}}].

\bibitem{electron-liquid}
G.~Giuliani and G.~Vignale, \emph{{Quantum Theory of the Electron Liquid}}.
\newblock Cambridge University Press, 2005.

\bibitem{he-huang}
L.~He and X.-G. Huang, \emph{{Nonperturbative Effects on the Ferromagnetic
  Transition in Repulsive Fermi Gases}},
  \href{http://dx.doi.org/10.1103/PhysRevA.85.043624}{\emph{Phys. Rev.} {\bf
  A85} (2012) 043624}, [\href{http://arxiv.org/abs/1106.1345}{{\tt
  1106.1345}}].

\bibitem{montse}
M.~Casas, C.~Esebbag, A.~Extremera, J.~M. Getino, M.~de~Llano, A.~Plastino
  et~al., \emph{Cooper pairing in a soluble one-dimensional many-fermion
  model}, \href{http://dx.doi.org/10.1103/PhysRevA.44.4915}{\emph{Phys. Rev. A}
  {\bf 44} (1991) 4915--4922}.

\bibitem{quick}
R.~M. Quick, C.~Esebbag and M.~de~Llano, \emph{{BCS} theory tested in an
  exactly solvable fermion fluid},
  \href{http://dx.doi.org/10.1103/PhysRevB.47.11512}{\emph{Phys. Rev. B} {\bf
  47} (1993) 11512--11514}.

\bibitem{strinati}
M.~Marini, F.~Pistolesi and G.~C. Strinati, \emph{{Evolution from BCS
  Superconductivity to Bose Condensation: Analytic Results for the crossover in
  three dimensions}}, {\emph{Eur. Phys. J.} {\bf 1} (1998) 151}.

\bibitem{fw}
A.~L. Fetter and J.~D. Walecka, \emph{Quantum theory of many-particle systems}.
\newblock Dover, 2012.

\bibitem{taka}
M.~Takahashi, \emph{{Many-Body Problem of Attractive Fermions with Arbitrary
  Spin in One Dimension}},
  \href{http://dx.doi.org/10.1143/PTP.44.899}{\emph{Progress of Theoretical
  Physics} {\bf 44} (1970) 899--904}.

\bibitem{sb-book}
L.~Samaj and Z.~Bajnok, \emph{{Introduction to the statistical physics of
  integrable many-body systems}}.
\newblock Cambridge University Press, 2013.

\bibitem{zhou-exact}
L.~Zhou, C.~yu~Xu and Y.~li~Ma, \emph{{Exact studies of ground and excited
  states of one-dimensional $\delta$-interacting Fermi gases in the BCS-BEC
  crossover}},
  \href{http://dx.doi.org/10.1088/1742-5468/2012/03/l03002}{\emph{Journal of
  Statistical Mechanics: Theory and Experiment} {\bf 2012} (2012) L03002}.

\bibitem{frz}
J.~N. Fuchs, A.~Recati and W.~Zwerger, \emph{{Exactly Solvable Model of the
  BCS-BEC Crossover}},
  \href{http://dx.doi.org/10.1103/PhysRevLett.93.090408}{\emph{Phys. Rev.
  Lett.} {\bf 93} (2004) 090408}.

\bibitem{guan-multi}
X.-W. Guan, Z.-Q. Ma and B.~Wilson, \emph{One-dimensional multicomponent
  fermions with $delta$-function interaction in strong- and weak-coupling
  limits: $\kappa$-component {F}ermi gas},
  \href{http://dx.doi.org/10.1103/PhysRevA.85.033633}{\emph{Phys. Rev. A} {\bf
  85} (2012) 033633}.

\bibitem{lang}
G.~Lang, \emph{Correlations in low-dimensional quantum gases}.
\newblock Springer--Verlag, 2018.

\bibitem{mr-ll}
M.~Mari\~no and T.~Reis, \emph{{Exact perturbative results for the
  Lieb--Liniger and Gaudin--Yang models}},
  \href{http://arxiv.org/abs/1905.09569}{{\tt 1905.09569}}.

\bibitem{ksv}
I.~Kostov, D.~Serban and D.~Volin, \emph{{Functional BES equation}},
  \href{http://dx.doi.org/10.1088/1126-6708/2008/08/101}{\emph{JHEP} {\bf 08}
  (2008) 101}, [\href{http://arxiv.org/abs/0801.2542}{{\tt 0801.2542}}].

\bibitem{fnw1}
P.~Forgacs, F.~Niedermayer and P.~Weisz, \emph{{The Exact mass gap of the
  Gross-Neveu model. 1. The Thermodynamic Bethe ansatz}},
  \href{http://dx.doi.org/10.1016/0550-3213(91)90044-X}{\emph{Nucl. Phys.} {\bf
  B367} (1991) 123--143}.

\bibitem{bbbkp}
Z.~Bajnok, J.~Balog, B.~Basso, G.~P. Korchemsky and L.~Palla, \emph{{Scaling
  function in AdS/CFT from the O(6) sigma model}},
  \href{http://dx.doi.org/10.1016/j.nuclphysb.2008.11.023}{\emph{Nucl. Phys.}
  {\bf B811} (2009) 438--462}, [\href{http://arxiv.org/abs/0809.4952}{{\tt
  0809.4952}}].

\bibitem{hmn}
P.~Hasenfratz, M.~Maggiore and F.~Niedermayer, \emph{{The Exact mass gap of the
  O(3) and O(4) nonlinear sigma models in d = 2}},
  \href{http://dx.doi.org/10.1016/0370-2693(90)90685-Y}{\emph{Phys. Lett.} {\bf
  B245} (1990) 522--528}.

\bibitem{hn}
P.~Hasenfratz and F.~Niedermayer, \emph{{The Exact mass gap of the O(N) sigma
  model for arbitrary $N \ge 3$ in $d = 2$}},
  \href{http://dx.doi.org/10.1016/0370-2693(90)90686-Z}{\emph{Phys. Lett.} {\bf
  B245} (1990) 529--532}.

\bibitem{msw}
M.~Mari\~no, R.~Schiappa and M.~Weiss, \emph{{Nonperturbative effects and the
  large-order behavior of matrix models and topological strings}},
  \href{http://dx.doi.org/10.4310/CNTP.2008.v2.n2.a3}{\emph{Commun. Num. Theor.
  Phys.} {\bf 2} (2008) 349--419}, [\href{http://arxiv.org/abs/0711.1954}{{\tt
  0711.1954}}].

\bibitem{bender-book}
C.~M. Bender and S.~A. Orszag, \emph{{Advanced mathematical methods for
  scientists and engineers}}.
\newblock Springer-Verlag, 1999.

\bibitem{mmnp}
M.~Mari\~no, \emph{{Nonperturbative effects and nonperturbative definitions in
  matrix models and topological strings}},
  \href{http://dx.doi.org/10.1088/1126-6708/2008/12/114}{\emph{JHEP} {\bf 0812}
  (2008) 114}, [\href{http://arxiv.org/abs/0805.3033}{{\tt 0805.3033}}].

\bibitem{rossi}
R.~Rossi, T.~Ohgoe, K.~Van~Houcke and F.~Werner, \emph{{Resummation of
  diagrammatic series with zero convergence radius for strongly correlated
  fermions}},
  \href{http://dx.doi.org/10.1103/PhysRevLett.121.130405}{\emph{Phys. Rev.
  Lett.} {\bf 121} (2018) 130405}, [\href{http://arxiv.org/abs/1802.07717}{{\tt
  1802.07717}}].

\bibitem{fkw}
V.~A. Fateev, V.~A. Kazakov and P.~B. Wiegmann, \emph{{Principal chiral field
  at large N}},
  \href{http://dx.doi.org/10.1016/0550-3213(94)90405-7}{\emph{Nucl. Phys.} {\bf
  B424} (1994) 505--520}, [\href{http://arxiv.org/abs/hep-th/9403099}{{\tt
  hep-th/9403099}}].

\bibitem{as}
A.~Altland and B.~D. Simons, \emph{Condensed matter field theory}.
\newblock Cambridge University Press, 2010.

\bibitem{mis-ov}
I.~A. Misurkin and A.~A. Ovchinnikov, \emph{{Analytic properties of the
  ground-state energy in the one-dimensional Hubbard model}},
  \href{http://dx.doi.org/10.1007/BF01028672}{\emph{Theoretical and
  Mathematical Physics} {\bf 11} (1972) 393--394}.

\bibitem{metzner}
W.~Metzner and D.~Vollhardt, \emph{Ground-state energy of the d=1,2,3
  dimensional {H}ubbard model in the weak-coupling limit},
  \href{http://dx.doi.org/10.1103/PhysRevB.39.4462}{\emph{Phys. Rev. B} {\bf
  39} (1989) 4462--4466}.

\bibitem{marsiglio}
F.~Marsiglio, \emph{{Evaluation of the BCS approximation for the attractive
  Hubbard model in one dimension}},
  \href{http://dx.doi.org/10.1103/PhysRevB.55.575}{\emph{Phys. Rev. B} {\bf 55}
  (1997) 575--581}.

\bibitem{bishop}
R.~Bishop, \emph{{Ground-state energy of a dilute Fermi gas}},
  \href{http://dx.doi.org/https://doi.org/10.1016/0003-4916(73)90411-9}{\emph{Annals
  Phys.} {\bf 77} (1973) 106--138}.

\bibitem{hf-effective}
H.~W. Hammer and R.~J. Furnstahl, \emph{{Effective field theory for dilute
  Fermi systems}},
  \href{http://dx.doi.org/10.1016/S0375-9474(00)00325-0}{\emph{Nucl. Phys.}
  {\bf A678} (2000) 277--294},
  [\href{http://arxiv.org/abs/nucl-th/0004043}{{\tt nucl-th/0004043}}].

\bibitem{fourth-order}
C.~Wellenhofer, C.~Drischler and A.~Schwenk, \emph{{Dilute Fermi gas at fourth
  order in effective field theory}},
  \href{http://arxiv.org/abs/1812.08444}{{\tt 1812.08444}}.

\bibitem{he}
L.~He, \emph{{Interaction energy and itinerant ferromagnetism in a strongly
  interacting Fermi gas in the absence of molecule formation}},
  \href{http://dx.doi.org/10.1103/PhysRevA.90.053633}{\emph{Phys. Rev.} {\bf
  A90} (2014) 053633}, [\href{http://arxiv.org/abs/1405.5242}{{\tt
  1405.5242}}].

\bibitem{pitaevski-rev}
S.~Giorgini, L.~P. Pitaevskii and S.~Stringari, \emph{{Theory of ultracold
  atomic Fermi gases}},
  \href{http://dx.doi.org/10.1103/RevModPhys.80.1215}{\emph{Rev. Mod. Phys.}
  {\bf 80} (2008) 1215--1274}, [\href{http://arxiv.org/abs/0706.3360}{{\tt
  0706.3360}}].

\bibitem{gross-neveu}
D.~J. Gross and A.~Neveu, \emph{{Dynamical Symmetry Breaking in Asymptotically
  Free Field Theories}},
  \href{http://dx.doi.org/10.1103/PhysRevD.10.3235}{\emph{Phys. Rev.} {\bf D10}
  (1974) 3235}.

\bibitem{fnw2}
P.~Forgacs, F.~Niedermayer and P.~Weisz, \emph{{The Exact mass gap of the
  Gross-Neveu model. 2. The 1/N expansion}},
  \href{http://dx.doi.org/10.1016/0550-3213(91)90045-Y}{\emph{Nucl. Phys.} {\bf
  B367} (1991) 144--157}.

\bibitem{montecarlo}
F.~Simkovic and E.~Kozik, \emph{{Determinant Monte Carlo for irreducible
  Feynman diagrams in the strongly correlated regime}},
  \href{http://arxiv.org/abs/1712.10001}{{\tt 1712.10001}}.

\bibitem{peris}
S.~Peris, \emph{{UV renormalons in QCD and their phenomenological
  implications}},
  \href{http://dx.doi.org/10.1016/S0920-5632(97)01086-4}{\emph{Nucl. Phys.
  Proc. Suppl.} {\bf 64} (1998) 344--349},
  [\href{http://arxiv.org/abs/hep-ph/9709488}{{\tt hep-ph/9709488}}].

\bibitem{nishida-son}
Y.~Nishida and D.~T. Son, \emph{{Fermi gas near unitarity around four and two
  spatial dimensions}},
  \href{http://dx.doi.org/10.1103/PhysRevA.75.063617}{\emph{Phys. Rev.} {\bf
  A75} (2007) 063617}, [\href{http://arxiv.org/abs/cond-mat/0607835}{{\tt
  cond-mat/0607835}}].

\bibitem{argyres-unsal}
P.~Argyres and M.~Unsal, \emph{{A semiclassical realization of infrared
  renormalons}},
  \href{http://dx.doi.org/10.1103/PhysRevLett.109.121601}{\emph{Phys. Rev.
  Lett.} {\bf 109} (2012) 121601}, [\href{http://arxiv.org/abs/1204.1661}{{\tt
  1204.1661}}].

\bibitem{a-unsal-long}
P.~C. Argyres and M.~Unsal, \emph{{The semi-classical expansion and resurgence
  in gauge theories: new perturbative, instanton, bion, and renormalon
  effects}}, \href{http://dx.doi.org/10.1007/JHEP08(2012)063}{\emph{JHEP} {\bf
  08} (2012) 063}, [\href{http://arxiv.org/abs/1206.1890}{{\tt 1206.1890}}].

\bibitem{dunne-unsal-cpn}
G.~V. Dunne and M.~Unsal, \emph{{Resurgence and Trans-series in Quantum Field
  Theory: The CP(N-1) Model}},
  \href{http://dx.doi.org/10.1007/JHEP11(2012)170}{\emph{JHEP} {\bf 11} (2012)
  170}, [\href{http://arxiv.org/abs/1210.2423}{{\tt 1210.2423}}].

\bibitem{cherman-dorigoni-dunne-unsal}
A.~Cherman, D.~Dorigoni, G.~V. Dunne and M.~Unsal, \emph{{Resurgence in Quantum
  Field Theory: Nonperturbative Effects in the Principal Chiral Model}},
  \href{http://dx.doi.org/10.1103/PhysRevLett.112.021601}{\emph{Phys. Rev.
  Lett.} {\bf 112} (2014) 021601}, [\href{http://arxiv.org/abs/1308.0127}{{\tt
  1308.0127}}].

\bibitem{cdu}
A.~Cherman, D.~Dorigoni and M.~Unsal, \emph{{Decoding perturbation theory using
  resurgence: Stokes phenomena, new saddle points and Lefschetz thimbles}},
  \href{http://dx.doi.org/10.1007/JHEP10(2015)056}{\emph{JHEP} {\bf 10} (2015)
  056}, [\href{http://arxiv.org/abs/1403.1277}{{\tt 1403.1277}}].

\bibitem{misumi-1}
T.~Misumi, M.~Nitta and N.~Sakai, \emph{{Classifying bions in Grassmann sigma
  models and non-Abelian gauge theories by D-branes}},
  \href{http://dx.doi.org/10.1093/ptep/ptv009}{\emph{PTEP} {\bf 2015} (2015)
  033B02}, [\href{http://arxiv.org/abs/1409.3444}{{\tt 1409.3444}}].

\bibitem{du-on}
G.~V. Dunne and M.~Unsal, \emph{{Resurgence and Dynamics of O(N) and
  Grassmannian Sigma Models}},
  \href{http://dx.doi.org/10.1007/JHEP09(2015)199}{\emph{JHEP} {\bf 09} (2015)
  199}, [\href{http://arxiv.org/abs/1505.07803}{{\tt 1505.07803}}].

\bibitem{misumi-2}
T.~Fujimori, S.~Kamata, T.~Misumi, M.~Nitta and N.~Sakai, \emph{{Bion
  non-perturbative contributions versus infrared renormalons in two-dimensional
  $\mathbb C P^{N-1}$ models}},
  \href{http://dx.doi.org/10.1007/JHEP02(2019)190}{\emph{JHEP} {\bf 02} (2019)
  190}, [\href{http://arxiv.org/abs/1810.03768}{{\tt 1810.03768}}].

\bibitem{iw-pol}
T.~Iida and M.~Wadati, \emph{Exact analysis of a $delta$-function spin-1/2
  attractive {F}ermi gas with arbitrary polarization},
  \href{http://dx.doi.org/10.1088/1742-5468/2007/06/p06011}{\emph{J. Stat.
  Mech. Theory Exp.} {\bf 2007} (2007) P06011--P06011}.

\bibitem{mr-ift}
M.~Mari\~no and T.~Reis, \emph{{Renormalons in integrable field theories}},
  \href{http://arxiv.org/abs/1909.12134}{{\tt 1909.12134}}.

\end{thebibliography}\endgroup

\end{document}